\documentclass[aps,twocolumn,prd,superscriptaddress]{revtex4-1}

\usepackage[utf8]{inputenc}

\usepackage{mathtools}
\usepackage{amsfonts}
\usepackage{mathrsfs}
\usepackage{bbm}
\usepackage{slashed}

\usepackage{graphicx}
\usepackage{color}
\usepackage{array}

\usepackage{placeins}
\usepackage{booktabs}
\usepackage{makecell}
\usepackage{subcaption}
\usepackage{floatrow}

\usepackage{xspace}
\usepackage{siunitx}
\usepackage{hyperref}

\usepackage{xifthen}

\setkeys{Gin}{width=0.48\textwidth}
\newcommand{\tr}{{\text{tr}}}

\def\eq#1{\eqref{#1}}
\newcommand{\eqrange}[2]{\eqref{#1}~--~\eqref{#2}}
\def\eqs#1{\eqref{#1}}
\def\Fig#1{Fig.~\ref{#1}}
\def\Figs#1#2{Fig.~\ref{#1} and \ref{#2}}
\def\Figrange#1#2{Fig. \ref{#1}~--~\ref{#2}}
\def\Tab#1{Tab.~\ref{#1}}
\def\Sec#1{Sec.~\ref{#1}}
\def\Section#1{Section~\ref{#1}}
\def\Secs#1#2{Sec.~\ref{#1} and \ref{#2}}
\def\App#1{App.~\ref{#1}}
\def\Apps#1#2{App.~\ref{#1} and \ref{#2}}

\newcommand*{\eg}{e.g.\@\xspace}
\newcommand*{\ie}{i.e.\@\xspace}

\newcommand*{\cf}{cf.\@\xspace}

\newcommand{\sumint}{\int\hspace{-4.8mm}\sum}
\DeclareMathOperator{\dd}{d}
\newcommand{\psym}{\bar{p}}
\newcommand{\perm}{\text{perm.}}
\newcommand{\imag}{\text{i}}
\newcommand{\colvec}[2]{\begin{pmatrix} #1 \\ #2 \end{pmatrix}}
\newcommand{\pvecsq}{\vec{p}^{\,2}}
\newcommand{\twopiT}{2\pi\, T}
\newcommand{\fourpiT}{4\pi\, T}
\newcommand{\eqnewline}{\nonumber\\[1.5ex]}
\newcommand{\momarg}[1]{\ifthenelse{\isempty{#1}}{}{\left( #1 \right)}}
\newcommand{\ltpre}[1][0pt]{\mathrel{\raisebox{#1}{{\tiny$\parallel$}}}}
\newcommand{\lt}{{\ltpre[1.2pt]}}
\newcommand{\pL}[2][]{\Pi^{\lt}_{#2}\momarg{#1}}
\newcommand{\pT}[2][]{\Pi^{\bot}_{#2}\momarg{#1}}
\newcommand{\pE}[2][]{\Pi^{\text{E}}_{#2}\momarg{#1}}
\newcommand{\pM}[2][]{\Pi^{\text{M}}_{#2}\momarg{#1}}
\newcommand{\GE}{G^{\text{E}}_T}
\newcommand{\Zc}{Z_c}
\newcommand{\ZA}{Z_A}
\newcommand{\ZAM}{\ZA^\text{M}}
\newcommand{\ZAE}{\ZA^\text{E}}
\newcommand{\alphaAcc}{\alpha_{\bar{c}cA}}
\newcommand{\lAcc}{\lambda_{\bar{c}cA}}
\newcommand{\lAccM}{\lambda^\text{M}_{\bar{c}cA}}

\newcommand{\alphaAAA}{\alpha_{A^3}}
\newcommand{\lAAA}{\lambda_{A^3}}
\newcommand{\lAAAM}{\lambda^\text{M}_{A^3}}
\newcommand{\lAAAE}{\lambda^\text{E}_{A^3}}
\newcommand{\alphaAAAA}{\alpha_{A^4}}
\newcommand{\lAAAA}{\lambda_{A^4}}
\newcommand{\lAAAAM}{\lambda^\text{M}_{A^4}}
\newcommand{\lAAAAE}{\lambda^\text{E}_{A^4}}
\newcommand{\LambdaTmin}{\Lambda_T^\text{min}} 
\newcommand{\myhfill}{\hfill\textcolor{white}{.}}
\newcommand{\CC}{C\nolinebreak\hspace{-.05em}\raisebox{.4ex}{\tiny\bf +}\nolinebreak\hspace{-.10em}\raisebox{.4ex}{\tiny\bf +}\@\xspace}

\newcommand{\EoM}{\text{\tiny{EoM}}}
\newcommand{\mgap}{m_{\text{\tiny gap}}}
\newcommand{\minText}{\text{\tiny{min}}}
\newcommand{\lqcd}{\Lambda_{\text{\tiny QCD}}}

\graphicspath{{./figures/}}

\newcommand{\gettitle}{Non-perturbative finite-temperature Yang-Mills theory}

\hypersetup{
	pdftitle={\gettitle},
	pdfauthor={Cyrol, Mitter, Pawlowski, Strodthoff},
	pdfkeywords={Yang-Mills theory} {Polyakov loop}
		{correlations functions} {temperature}
		{Debye mass} {confinement}  {deconfinement} 
		{functional renormalisation group}
		{background field} {order parameter},
	bookmarksopen=true,
	bookmarksopenlevel=2,
	bookmarksnumbered=true
}

\begin{document}

\title{\gettitle}

\author{Anton K. Cyrol}
\affiliation{Institut f\"ur Theoretische Physik,
	Universit\"at Heidelberg, Philosophenweg 16,
	69120 Heidelberg, Germany
}

\author{Mario Mitter}
\affiliation{Institute of Physics,
	NAWI Graz, University of Graz, Mozartgasse 14,
	8010 Graz, Austria
}

\author{Jan M. Pawlowski} 
\affiliation{Institut f\"ur Theoretische Physik,
	Universit\"at Heidelberg, Philosophenweg 16,
	69120 Heidelberg, Germany
}
\affiliation{ExtreMe Matter Institute EMMI,
	GSI, Planckstr. 1,
	64291 Darmstadt, Germany
}

\author{Nils Strodthoff}
\affiliation{Nuclear Science Division,
	Lawrence Berkeley National Laboratory,
	Berkeley, CA 94720, USA
}

\begin{abstract}
	We present non-perturbative correlation functions in Landau-gauge
	Yang-Mills theory at finite temperature. The results are obtained
	from the functional renormalisation group within a self-consistent
	approximation scheme. In particular, we compute the magnetic and electric
	components of the gluon propagator, and the three- and four-gluon
	vertices. We also show the ghost propagator and the
	ghost-gluon vertex at finite temperature. Our results for the propagators
	are confronted with lattice simulations and our Debye mass is 
	compared to hard thermal loop perturbation
	theory.
	\vspace*{52pt}
\end{abstract}

\maketitle

\section{Introduction}
\label{sec:Introduction}
Understanding the phase structure of Quantum Chromodynamics still
poses a major challenge. On the theoretical side, its
strongly-correlated nature hampers the progress of first-principles
approaches in particular at high densities.  While lattice simulations
struggle with the sign problem, functional methods have to address the
resonant interaction structure, which requires particularly advanced
truncations of the corresponding generating functionals. The
tremendous progress in functional approaches to QCD has recently led
to a shift from qualitative bottom-up towards quantitative top-down
approaches
\cite{Mitter:2014wpa,Cyrol:2016tym,Cyrol:2017ewj,Binosi:2014aea,Williams:2015cvx,Huber:2016tvc},
see also \cite{Berges:2000ew,Pawlowski:2005xe,Gies:2006wv,
  Schaefer:2006sr,Rosten:2010vm,Braun:2011pp,vonSmekal:2012vx,Alkofer:2000wg,
  Roberts:2000aa,Fischer:2006ub,Fischer:2003rp,Fischer:2008uz,Binosi:2009qm,
  Maas:2011se, Boucaud:2011ug,Eichmann:2016yit} for reviews.  In
particular, the functional renormalisation group (FRG) is a
first-principles method that allows for quantitative computations of
the generating functional of QCD.  Recently, the functional QCD (fQCD)
collaboration \cite{fQCD:2016-10} has established a comprehensive
framework encompassing both, top-down
\cite{Mitter:2014wpa,Cyrol:2016tym,Cyrol:2017ewj} and bottom-up
\cite{Haas:2013qwp,Herbst:2013ufa,Braun:2014ata,Rennecke:2015eba,Fu:2015naa,Heller:2015box,Fu:2016tey}
approaches within the FRG framework.

In this work, we focus on thermal correlation functions of the pure
gauge sector of QCD. Quantitative control over Yang-Mills theory
at zero as well as finite temperature is a pivotal prerequisite for
predictive investigations of the QCD phase structure with functional
methods. While vacuum Yang-Mills correlation functions have been
studied intensively in the past two decades
\cite{vonSmekal:1997ohs,Zwanziger:2001kw,Lerche:2002ep,Fischer:2002eq,Gies:2002af,Pawlowski:2003hq,Fischer:2004uk,
  Alkofer:2004it,Fischer:2006vf,Huber:2007kc,Alkofer:2008jy,Aguilar:2008xm,Boucaud:2008ky,Fischer:2009tn,Tissier:2010ts,Tissier:2011ey,
  Huber:2012kd,Aguilar:2013xqa,Pelaez:2013cpa,Quandt:2013wna,Aguilar:2013vaa,Blum:2014gna,Eichmann:2014xya,
  Gracey:2014mpa,Gracey:2014ola,Binosi:2014kka,Cyrol:2014kca,Huber:2014isa,Cyrol:2016tym,Athenodorou:2016oyh,Reinosa:2017qtf},
results for the finite-temperature correlation functions are scarce,
see
\cite{Maas:2004se,Fister:2011uw,Quandt:2015aaa,Reinosa:2013twa,Reinosa:2016iml,Cucchieri:2007ta,Fischer:2010fx,Aouane:2011fv,Maas:2011ez,Cucchieri:2012gb,Silva:2013maa}
for propagator studies. For the vertices, the situation is even less
satisfactory and only exploratory studies exist
\cite{Fister:2014bpa,Huber:2016xbs}.

The main goal of this study is to get quantitative access to the
finite-temperature $1$PI $n$-point functions of Yang-Mills theory.
These correlators contain all the information about the
observables. For example, the resulting propagators and vertices can
be used to investigate the center-symmetry phase transition in terms
of the Polyakov-loop potential, see \eg
\cite{Braun:2007bx,Marhauser:2008fz,Braun:2009gm,Fukushima:2012qa,Fister:2013bh,Herbst:2015ona,Braun:2010cy,Dumitru:2012fw,Reinosa:2014zta,Reinosa:2015gxn,Quandt:2016ykm,Fukushima:2017csk}.
Furthermore, the Debye mass, which has been studied intensively on the
lattice \cite{Heller:1997nqa,Datta:1999yu,Cucchieri:2001tw,Nakamura:2003pu}
and hard thermal loop perturbation theory \cite{Andersen:2009tc,Andersen:2010ct}
as well as with other thermal QCD approaches \cite{Arnold:1995bh,Braaten:1995ju,Braaten:1995jr},
can be extracted from the gluon propagator.
Additionally, the
correlators allow for the extraction of spectral functions and the
calculation of the shear viscosity
\cite{Haas:2013hpa,Christiansen:2014ypa}.

To calculate the $1$PI $n$-point functions, we perform a systematic
vertex expansion of the effective action with the aim of quantitative
precision, controlled by apparent convergence.  The zero-temperature
baseline for this calculation is provided by \cite{Cyrol:2016tym}, a
recent FRG study, which incorporates all tensor structures present at
the classical level in a self-consistent truncation scheme. Here, we
generalise this truncation to finite temperature, which includes the
splitting of the correlation functions into electric and magnetic
components.  In particular, we provide results for the electric and
magnetic gluon propagators as well as the electric and magnetic components
of the three- and four-gluon vertices. For the propagators, we compare
extensively to results obtained in lattice simulations.
We use the Debye screening mass to determine a lower bound for the temperature range
in which hard thermal loop perturbation theory can be applied straightforwardly.
Furthermore, the finite-temperature behaviour of
the ghost-induced zero crossing of the three-gluon vertex is
investigated.  The comprehensive truncation brings along new technical
challenges whose solutions are discussed. In summary, this work
provides a major step towards investigations of the QCD phase
structure from first principles within the functional methods.

This paper is organised as follows.
In \Sec{sec:Setup} we discuss the finite-temperature vertex expansion,
order parameters, and the Debye screening mass.
\Section{sec:Method} deals with finite-temperature flows of gauge
theories.  We present our results in \Sec{sec:Results} and discuss
them in \Sec{sec:Discussion}.  Finally, we summarise our findings and
give an outlook in \Sec{sec:Conclusion}.  Technical details and
numerical checks are provided in the appendices.
In particular, we confirm regulator independence in \App{app:RegulatorIndependence}.

\section{Yang-Mills theory at \texorpdfstring{$\mathbf{T>0}$}{T>0}}
\label{sec:Setup}

We consider Euclidean Yang-Mills theory, whose classical action in general covariant 
gauges is given by
\begin{align}
	S= \frac{1}{4} \int_x\, F_{\mu\nu}^a  F_{\mu\nu}^a
				+ \frac{1}{2\xi}\int_x\,(\partial_\mu A^a_\mu)^2
				- \int_x\,\bar c^a \partial_\mu D^{ab}_\mu c^b\,.
	\label{eq:ClassicalAction}
\end{align}
Here, $A,\, c$ and $\bar c$ denote the gluon, ghost and antighost fields
and $\int_x=\int \dd^4 x\,$. The gauge fixing parameter $\xi$ is taken to 
zero in Landau gauge, $\xi\rightarrow 0\,$. The field strength tensor and 
adjoint covariant derivative are given by
\begin{align}
	F^a_{\mu\nu} &=  \partial_\mu A^a_\nu-\partial_\nu A^a_\mu+
		g f^{abc}A_\mu^b A_\nu^c\, , \eqnewline
	D^{ab}_{\mu} &= \delta^{ab}\partial_\mu - g f^{abc} A^c_\mu\,,
\end{align}
where $f^{abc}$ are the structure constants of the Lie algebra.
Our notation largely follows earlier works within the fQCD
collaboration \cite{fQCD:2016-10}, and we refer to 
\cite{Mitter:2014wpa,Braun:2014ata,Cyrol:2016tym,Cyrol:2017ewj} for further
details.

\subsection{Finite-temperature vertex expansion}
\label{sec:Setup:VertexExpansion}

Functional approaches require an approximation of the corresponding
generating functional. We use a vertex expansion about the vanishing
expectation values of the gluon and ghosts fields, $A_\mu=0$ and
$c=\bar c=0\,$. These field values are solutions of the equations of
motion and constitute the vacuum at vanishing temperature. The
intricacies at finite temperature are discussed in more detail in the
next \Sec{sec:Setup:ExpansionPoint}. In the vertex expansion, the
effective action is written as a sum over powers of the fields,
\begin{align}
	\label{eq:VertexExpansion}
	\Gamma[\Phi] = \sum_n  \frac{1}{n!}\,
	\sumint_{\{p_i\}}\Gamma^{(n)}(p_1,\dots,p_n)
		\;\Phi (p_1)\cdots\Phi (p_n)\,,
\end{align}
where $\Phi=(A_\mu,\,\bar{c},\,c)$ is a superfield and
momentum conservation implies $\sum_i p_i=0\,$.
The expansion coefficients in \eq{eq:VertexExpansion} are the $1$PI $n$-point
functions that are in field components given by
\begin{align}
	\label{eq:Vertices}\Gamma^{(n)}_{\Phi_{i_1}\cdots\Phi_{i_n}}[\Phi]
		=\frac{\delta^n \Gamma_k[\Phi]}{\delta \Phi_{i_n}\cdots \delta \Phi_{i_1}}\,.
\end{align}
The correlation functions are expanded in terms of
basis tensors ${\cal T}^{(i)}$ and dressing functions $\lambda^{(i)}\,$,
\begin{align}
	\label{eq:ProperVertices}
	\Gamma^{(n)}_{\Phi_{i_1}\dots\Phi_{i_n}} &= \sum\limits_{i} 
		\lambda^{(i)}_{\Phi_{i_1}\dots\Phi_{i_n}}{\cal T}^{(i)}_{\Phi_{i_1}\dots\Phi_{i_n}}\,.
\end{align}

\begin{figure}
	\begin{tabular}{m{0.23\textwidth}m{0.23\textwidth}m{0.23\textwidth}m{0.23\textwidth}m{0\textwidth}}
		\centering\includegraphics[width=0.16\textwidth]{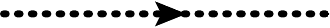} &
		\centering\includegraphics[width=0.15\textwidth]{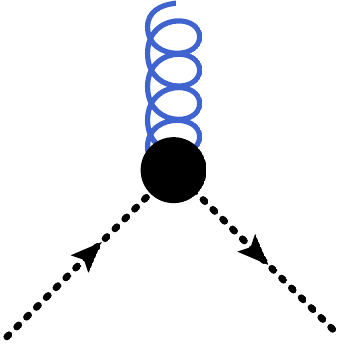} &
		\centering\includegraphics[width=0.16\textwidth]{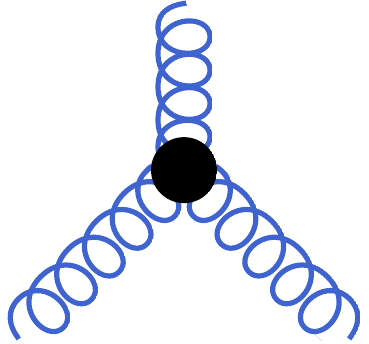} &
		\centering\includegraphics[width=0.16\textwidth]{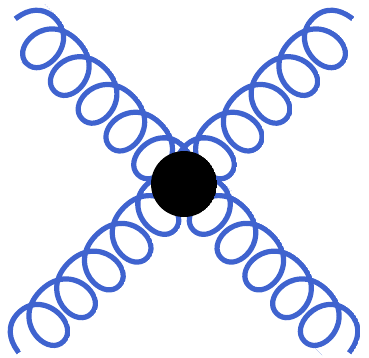} & \\ 
		\centering $1/\Zc(\psym)$ & \centering $\lAccM(\psym)$ & \centering  $\lAAAM(\psym)$ & \centering $\lAAAAM(\psym)$  &\\
		&&&&\\
		\centering\includegraphics[width=0.16\textwidth]{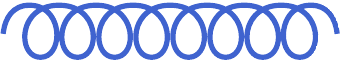} &
		\centering\includegraphics[width=0.16\textwidth]{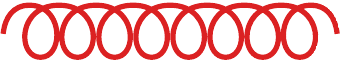} &
		\centering\includegraphics[width=0.16\textwidth]{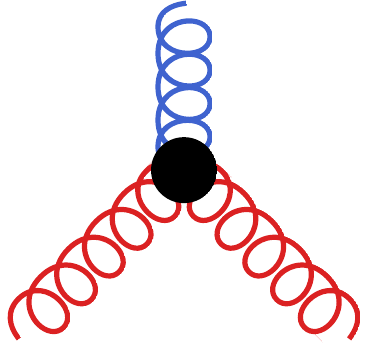} & 
		\centering\includegraphics[width=0.16\textwidth]{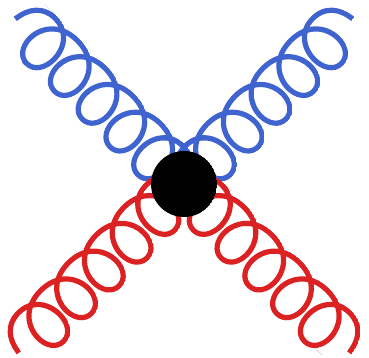}  &\\ 
		\centering $1/\ZAM(\psym)$ & \centering $1/\ZAE(\psym)$ & \centering $\lAAAE(\psym)$ & \centering $\lAAAAE(\psym)$ &
	\end{tabular}
	\caption{
		Constituents of our vertex expansion.
		We use the classical tensors that are present in 
		the bare action and attach magnetic (blue) and electric (red) 
		projection operators to the gluon legs.
		Missing combinations, \eg vertices with one electric leg, vanish 
		if the Matsubara modes are set to zero and are not computed in
		our truncation.\myhfill
	}
	\label{fig:Truncation}
\end{figure}

At finite temperature, the vacuum $O(4)$-symmetry is replaced by
$\mathbb{Z}_2 \otimes O(3)\,$. This reduced symmetry implies a
difference between the magnetic and electric components, which
correspond to the directions that are transverse and longitudinal with
respect to the heat bath. Starting from the longitudinal and
transverse vacuum projection operators,
\begin{align}
	\pL[p]{\mu\nu} &= \frac{p_\mu p_\nu}{p^2}\, ,\eqnewline
	\pT[p]{\mu\nu} &= \delta_{\mu\nu}-\pL[p]{\mu\nu}\,,
	\label{eq:VacuumProjectors}
\end{align}
we decompose four-vectors into
\begin{align}
	\label{eq:finiteTvector}
	p = \colvec{\omega_n}{\vec{p}} = \colvec{\twopiT\, n }{\vec{p}}\,,
\end{align}
where $n\in\mathbb{Z}$ are the discrete Matsubara modes and
$\omega_n=\twopiT\, n$ the corresponding frequencies. This leads to
the magnetic and electric projection operators at finite temperature,
\begin{align}
	\label{eq:finitetProjectors}
	\pM[p]{\mu\nu} &= (1-\delta_{0\mu})(1-\delta_{0\nu})
		\left(\delta_{\mu\nu}-\frac{p_\mu p_\nu}{\pvecsq}\right)\,,\eqnewline
	\pE[p]{\mu\nu} &= \pT[p]{\mu\nu}-\pM[p]{\mu\nu}\,.
\end{align}
A crucial consequence of the breaking of the vacuum $O(4)$-symmetry by
\eq{eq:finitetProjectors} is the splitting of the tensor structures
into electric and magnetic components.  In particular, the propagators
are given by
\begin{align}
	\label{eq:propagators}
	&[\Gamma^{(2)}_{AA}]^{ab}_{\mu\nu} (p) = \delta^{ab} \, p^2 
		\left[\ZAM(p) \; \pM[p]{\mu\nu} + \ZAE(p) \; \pE[p]{\mu\nu} \right]\,,\eqnewline
	&[\Gamma^{(2)}_{\bar{c}c}]^{ab}(p) \,\, = \delta^{ab} \, p^2\, \Zc(p)\,,
\end{align}
with dimensionless scalar dressing functions $1/\ZAM$ and $1/\ZAE$ for
the magnetic and electric components of the gluon propagator.  In the
case of the vertices, we take only the classical tensor structures
into account.  Similarly to the gluon propagator, we split their
dressings into electric and magnetic components. See
\Fig{fig:Truncation} for an illustration of the constituents of our
truncation and \Apps{app:TensorSplitting}{app:Degeneracy} for further
details.  As a consequence of the restriction to classical tensors
only, the tensor bases of the gluonic vertices are not complete, and
the projection of the tensor equations onto the scalar dressing
functions is not unique.  We use vacuum calculations to identify
uncertainties that stem from the projections in order to disentangle
them from finite-temperature effects, see \App{app:Projections} for
details.

Due to the breaking of $O(4)$-invariance, the scalar dressings are in
general functions of the Matsubara modes and spatial momenta, \eg
$Z(p) = Z(\omega_n^2,\,\pvecsq)$ for a generic wave function
renormalisation $Z$. However, the thermal contributions to the
correlation functions decay rapidly for spatial momenta and
frequencies with $p^2 \gtrsim (\twopiT)^2\,$.  Hence, 
the thermal correlation functions converge quickly towards
their $O(4)$-symmetric vacuum counterparts for these
momenta, see \eg \Figs{fig:MagneticGluonProp}{fig:ThreeGluonVertex}. 
Consequently, the
correlation functions exhibit an approximate $O(4)$-symmetry for all
higher Matsubara modes, and most of the finite temperature
effects are encoded in the zero mode at small spacial momenta
$\pvecsq \lesssim (\twopiT)^2\,$.  Therefore, the spatial momentum
dependence of the Matsubara zero modes can be used to obtain a very
good approximation of the full frequency and momentum dependence via
\begin{align}
	\label{eq:WaveApprox}
	Z(\omega_n^2,\,\pvecsq)\approx Z(0,\,\omega_n^2 + \pvecsq)\,, 
\end{align}
or in short $Z(p)=Z(0,\,p^2)\,$. 
In this work, we compute the zero modes of the propagators and
use \eq{eq:WaveApprox} to close the equations.
Within functional methods, this $O(4)$-symmetric approximation has
been found to be quantitatively reliable for gluon 
\cite{Fister:2011uw,Fister:Diss} as well as quark propagators \cite{Fischer:2010fx}.
This is confirmed by lattice studies that show a slight deviation 
of the $O(4)$ invariance only for the first Matsubara mode at temperatures
just below the critical temperature \cite{Ilgenfritz:2017kkp,Silva:2017feh}.

This pattern carries over to the scalar dressings of higher
order correlation functions $\lambda^{(n)}(p_1,\,\dots,\,p_n)\,$.
Analogously to the propagator dressings,
we base our computation on the zero modes
\begin{align}
	\label{eq:o4-approx}
	\lambda^{(n)}(\vec{p}_1,\,\dots,\,\vec{p}_i)
		=\lambda^{(n)}(\omega_{n_1}=0,\,\vec{p}_1,\,\dots,\,\omega_{n_n}=0,\,\vec{p}_n)\,.
\end{align} 
In contradistinction to the propagator dressings,
the zero modes of the vertex dressings $\lambda^{(n)}$ depend on
all $\vec{p}_i \cdot \vec{p}_j$ and not only $\pvecsq\,$.
However, the spatial momentum dependence of the
vertices is well described by a one-dimensional symmetric-point
approximation in four \cite{Cyrol:2016tym} as well as three 
dimensions \cite{Corell:Prep,Corell:MSc}, see also
\App{app:RegulatorIndependence}. This leads to
\begin{align}
	\label{eq:generalMomentumApproxzero}
	\lambda^{(n)}(\vec{p}_1,\,\dots,\,\vec{p}_n) \approx 
		\lambda^{(n)}\left(\bar p\right)\,, \qquad
		\psym^2 \equiv \frac{1}{n}\sum_{i=1}^n\pvecsq_i\,,
\end{align}
which allows to compute the flows of the zero modes of the vertex
dressings in a quantitatively reliable approximation, 
\cf \Fig{fig:RegulatorIndependence}. However, the flows of the zero modes
depend on the full frequency and spatial momentum dependence.
Analogously to the propagator dressings, 
we approximate the full momentum dependence 
with an $O(4)$-symmetric generalisation of \eq{eq:generalMomentumApproxzero},
\begin{align}
	\label{eq:generalMomentumApprox}
	\lambda^{(n)}(\omega_{n_1},\,\vec{p}_1,\,\dots,\,\omega_{n_n},\,\vec{p}_n) \approx 
		\lambda^{(n)}\left(\bar p\right)\,, 
\end{align}
where the symmetric momentum $\bar p$ is then given by
\begin{align}
	\qquad \psym^2 \equiv \frac{1}{n}\sum_{i=1}^n\left(\omega_{n_i}^2 + \pvecsq_i\right)\,.
\end{align}
In summary, we use two quantitatively reliable approximations for the dressing
functions: the approximate $O(4)$-invariance of all non-vanishing Matsubara modes,
which allows to use only information from the lowest Matsubara 
mode, and the well-tested symmetric point approximation. This truncation generalises 
the vacuum truncation used in \cite{Cyrol:2016tym}, 
see \App{app:VacuumLimit} for an explicit numerical check.

\subsection{Non-trivial vacuum and backgrounds}
\label{sec:Setup:ExpansionPoint}

As discussed in the last \Sec{sec:Setup:VertexExpansion}, we use a
vertex expansion about vanishing field expectation values $A_\mu=0$ and
$c=\bar c=0\,$. This necessitates a thorough discussion of
the implications of this choice, in particular for comparisons to
lattice results. We argue that such an expansion about vanishing background
fields, \ie Landau gauge, leads to correlation functions that
agree with the lattice correlators for temperatures
outside a small region around the phase transition. 
Furthermore, even near the phase transition, 
sizeable effects are only expected for correlation
functions that have electric gluon legs, the electric
gluon propagator being their most prominent representative. 
This becomes most evident by investigating the relation of
the physical solution of the equation of motion in non-vanishing 
gluon background fields and the Polyakov loop, 
the canonical order parameter of the confinement-deconfinement 
phase transition.  For the convenience of the reader, the first two parts 
briefly review corresponding relevant advances in non-perturbative 
functional approaches
\cite{Braun:2007bx,Marhauser:2008fz,Braun:2009gm,Spallek:MSc,Fukushima:2012qa,Fister:2013bh,Herbst:2015ona},
see \cite{Braun:2010cy,Dumitru:2012fw,
  Reinosa:2014zta,Reinosa:2015gxn,Quandt:2016ykm,Reinosa:2016iml,Fukushima:2017csk} for
further applications.

\subsubsection{Correlation functions}

To facilitate the discussion, we use the background extension of Landau gauge, 
\begin{align}
	\label{eq:bgauge} 
	(\partial_\mu -\imag g \bar A_\mu ) \,a_\mu =0\quad{\rm with}\quad  
	A_\mu = \bar A_\mu +a_\mu\,, 
\end{align}
called Landau-deWitt gauge. Here, $\bar A_\mu$ is a general background and
$a_\mu$ is the quantum fluctuation field. In this formulation, the
effective action is gauge invariant under background gauge
transformations, which allows for a simpler interpretation of physical
backgrounds as well as simpler technical implementations.
Besides being a functional of $\Phi =(a_\mu,\,\bar{c},\,c)\,$, 
the effective action depends now also on the background $\bar A\,$. 
Accordingly, the vertices $\Gamma^{(n)}[\bar A,\Phi]$
are correlation functions in the background
\begin{align}
	\label{eq:mean}
	\Phi=\langle \hat\Phi\rangle_{J(\bar A,\Phi)}\,.
\end{align} 
Here, we have introduced the background current, which satisfies $J(\bar A,\Phi)=\delta \Gamma/\delta\Phi\,$.
The correlation functions in the absence of external sources, $J(\bar A,\Phi)=0\,$,
are then given by $\Gamma^{(n)}[\bar A,\Phi_\EoM]\,$, where $\Phi_\EoM$ is a solution of the
equation of motion in the chosen background $\bar A\,$,
\begin{align}
	\label{eq:EoM}
	\left.\frac{\delta \Gamma[\bar A;\Phi]}{\delta\Phi}\right|_{\Phi_\EoM}=0\,.
\end{align}
In general, this conditions yields stationary points of the effective action.
In particular, the expansion point $(\bar A,\Phi)=0$ satisfies \eq{eq:EoM}, 
but does not necessarily single out the physical minimum.
In contrast, the physical correlators that correspond to scattering
amplitudes are obtained at the physical solution of
the equation of motion \eq{eq:EoM}, i.e. the minimum of the effective action
$(\bar A,\Phi_\minText[\bar A])\,$. This is also the
field value about which the vertex expansion is expected to be most stable and
converge most rapidly. Furthermore, only an expansion around the physical solution of 
the equation of motion allows for a direct comparison with correlation functions 
from lattice simulations, since the latter are measured on the physical ground state. 
In general, any other expansion point requires information about
higher correlation functions in order to evaluate $\Gamma^{(n)}[\bar A,\Phi_\minText]\,$.
In particular, in a vertex expansion with expansion point $(\bar A,\Phi)=0\,$, the inverse
propagator is given by
\begin{align}
	\Gamma^{(2)}[\bar A,\Phi_\minText] = \sum_{n} 
		\frac{1}{n!}\, \sumint_{\{p_i\}} \, & \Gamma^{(2+n)}[\bar A,0](p_1,\ldots,p_n) \eqnewline
		& \times \Phi_{\minText}(p_1)\cdots\Phi_{\minText}(p_n)\,,
	\label{eq:exp2}
\end{align}
where we suppressed external momentum arguments.
Therefore, we expect deviations between the correlation functions $\Gamma^{(n)}[0,0]\,$, computed in this work, and those
from lattice simulations. However, these differences are sizeable only
if the momentum scales of the solution $(\bar A,\Phi_\minText)\neq 0$ are
of the same order as $T_c\,$, the characteristic scale of the finite temperature Yang-Mills
system. Only in this case, the higher correlation functions would lead to noticeable
contributions in \eq{eq:exp2}. 

We can utilise the background field to achieve a technical simplification.
Since it is arbitrary, we can choose $\bar A= \bar A_\minText\equiv \langle \bar A\rangle$ such that
\begin{align}
	\label{eq:BEoM}  
		\Phi_\minText[\bar A_\minText]=0\,.
\end{align}
For this particular choice, the background carries all the non-trivial information about the ground state,
whereas the (classical) fluctuation field vanishes on the equation of motion. 
The physical correlators are then given by $\Gamma^{(n)}[\langle\bar A\rangle,0]\,$.
In particular, the inverse propagator \eq{eq:exp2} for the gluon is then given by
\begin{align}
	\Gamma^{(2)}_{AA}[\bar A,\Phi_\minText] = \Gamma^{(2)}_{AA}[\langle\bar A\rangle,0]\,.
	\label{eq:exp2_min}
\end{align}

Semi-perturbative studies in the Curci-Ferrari model for Yang-Mills
theory confirm that the background has large effects on the electric
propagator at temperatures close to the phase transition
\cite{Reinosa:2016iml}.  Furthermore, the calculation of
quantitatively correct values for the chiral phase transition
temperature as well as its observed coincidence with the
confinement-deconfinement crossover temperature require to take into
account such a non-trivial minimum \cite{Braun:2009gm}. Finally, such
a consistent treatment was also required for the description of the
Roberge-Weiss transition \cite{Braun:2009gm} as well as the study of
criticality in $SU(2)$ Yang-Mills theory
\cite{Marhauser:2008fz,Spallek:MSc}.

\begin{figure}
	\includegraphics[width=0.96\textwidth]{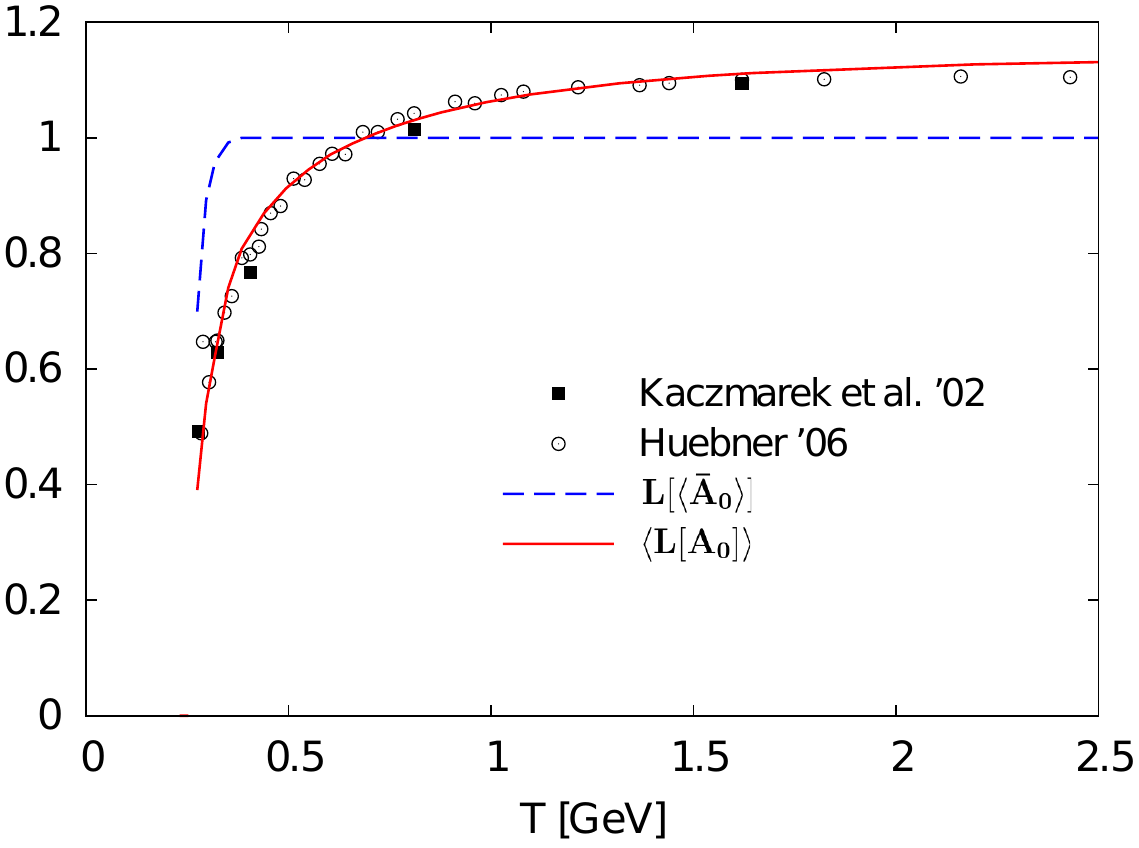}
	\caption{
		Expectation value of $\langle L[ A_0]\rangle$
		versus $L[\langle \bar A_0\rangle]$ from
		\cite{Herbst:2015ona}. Both observables are order parameters
		for the confinement-deconfinement phase
		transition. Moreover, $L[ \langle \bar A_0\rangle]=1$
		entails $\langle \bar A_0\rangle=0\,$.\myhfill}
	\label{fig:OrderParameters}
\end{figure}

\begin{figure*}
	\ffigbox{
		\begin{subfloatrow}
			\ffigbox{\includegraphics{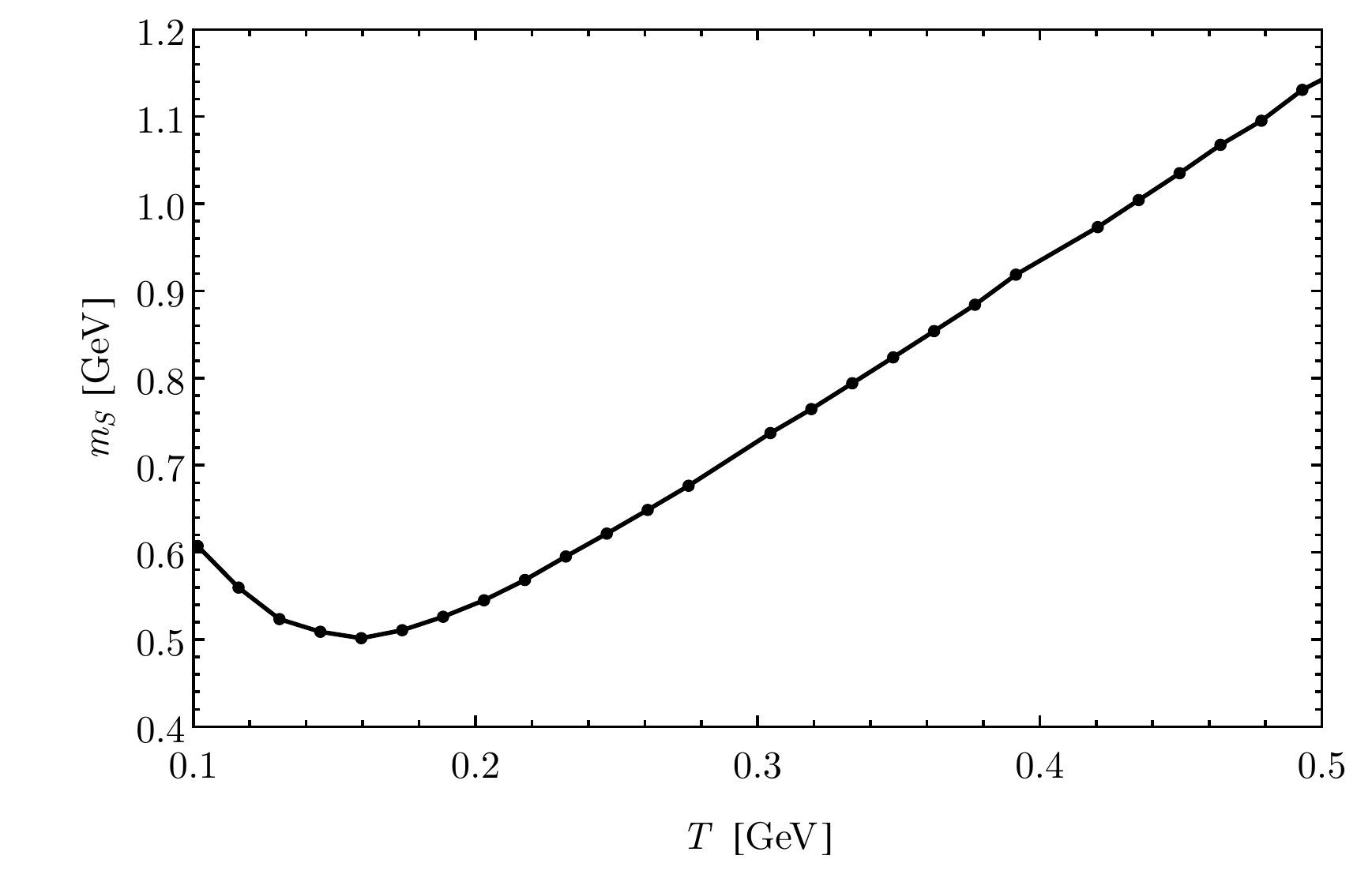}}{
				\caption{Screening mass $m_s$ in units of \si{GeV} at low
					temperatures.
					\myhfill}
				\label{fig:DebyeMass:LowT}
			}
			\ffigbox{\includegraphics{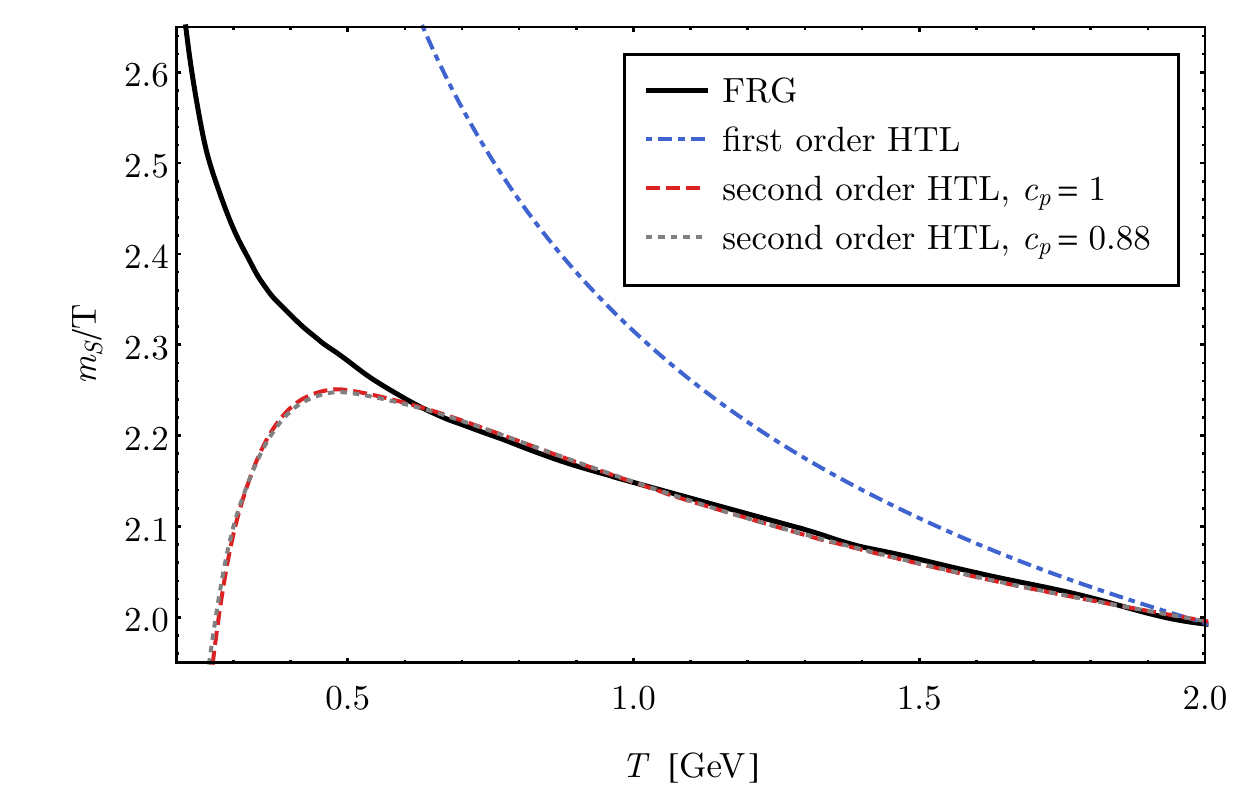}}{
				\caption{Dimensionless Debye screening  $m_s/T$
					mass at high temperatures
					in comparison with leading order perturbation theory
					\eq{eq:DebyeMass1} and the Arnold-Yaffe prescription \eq{eq:DebyeMass2}
					for accommodating beyond leading order effects \cite{Arnold:1995bh}.\myhfill}
				\label{fig:DebyeMass:HighT}
			} 
		\end{subfloatrow}
	}{
		\caption{
			Debye screening mass $m_s$, 
			see \Fig{fig:DebyeMassFit} for the fits of \eq{eq:DecayT} to $G_T(x)\,$.
		}
		\label{fig:DebyeMass}
	}
\end{figure*}

\subsubsection{Order parameters}

A further advantage of the background $\langle\bar A\rangle$ is 
its relation to the Polyakov loop~\cite{Braun:2007bx,Braun:2009gm,Fister:2013bh,Herbst:2015ona}, 
the standard order parameter of the confinement-deconfinement phase transition.
The traced Polyakov loop is expressed as a correlator of the temporal
gauge field with
\begin{align}
	&L[A_0]=\frac{1}{N_c}\tr\, P[A_0]\,,\eqnewline 
	&P[A_0]  = {\cal P} e^{\imag g \int_0^\beta \dd t \, A_0(t,\vec x)}\equiv e^{2 \pi \imag\,\varphi[A_0]}\,.
		\label{eq:Lphi}
\end{align}
Here, $ {\cal P}$ stands for path ordering, and the functional
$\varphi[A_0]$ is the gauge-covariant algebra element of the Polyakov
loop. It transforms as $\varphi \to U \varphi[A_0] U^\dagger$ under
time-periodic gauge transformation $U\in SU(N)\,$. This entails that
the eigenvalues of $\varphi[A_0]$ are gauge invariant, and
consequently the eigenvalues of its expectation value
$\bar \varphi\equiv\langle \varphi[A_0] \rangle$ are observables.
This expectation value, as well as $L[\langle \varphi\rangle]\,$, are
order parameters for the confinement-deconfinement phase transition.
In Polyakov gauge, the expectation value of the algebra element takes
the particularly simple form
\begin{align}
	\bar \varphi= \beta g \bar A_0\,,
\end{align}
for a given background $\bar A_0$. Due to background gauge invariance,
the eigenvalues of $\bar \varphi$ can therefore be calculated from the
eigenvalues of $\bar A_0$ in any background gauge. In particular, the
effective potential, $V[\bar A_0]$, of Landau-deWitt gauge carries
thus the full information about the eigenvalues of the expectation
value of $\varphi\,$.

In conclusion, the effective potential $V[\bar A_0]$ is an order
parameter potential for center symmetry.  The gauge invariant
observables, $\langle L[A_0]\rangle$ and $L[\langle\bar A_0\rangle]$,
or equivalently also $\langle\bar A_0\rangle$, serve as order
parameters for the confinement-deconfinement phase transition, see
\Fig{fig:OrderParameters}. Therefore, the vanishing of $\langle L[A_0]\rangle$ in
the confined phase relates to a non-vanishing value for
$\langle\bar A_0\rangle$. This has recently been demonstrated
explicitly in a self-consistent vertex expansion scheme, which has
been used for the first computation of $\langle L[A_0]\rangle$ within
functional methods \cite{Herbst:2015ona}.  Finally, the electric
propagator $\langle A_0(p) A_0(-p)\rangle$ is closely related to the
propagator of an order parameter field, and as such should show
critical properties, see \cite{Fister:2013bh}. Hence, we expect the
electric correlators to be affected most by the background field.

\subsubsection{Comparison to lattice simulations}

The previous discussion of the non-trivial $\bar A_0$ background and
its relation to the order parameter of the confinement-deconfinement
phase transition allows us to derive a theoretical estimate of the
temperature range, in which our present results potentially deviate
from the respective lattice results due to the different background
configurations. The first important piece of information is given by
the fact that the order parameter $L[\langle\bar A_0\rangle]$
approaches unity rapidly for temperatures above the phase transition
temperature, see \Fig{fig:OrderParameters}.  This in stark contrast to the
Polyakov loop $\langle L[ A_0]\rangle\,$, which is usually
calculated in lattice simulations. The latter reaches its asymptotic value
only for $T\gg T_c\,$, which can be understood from fluctuation
effects \cite{Herbst:2015ona}.  The fact that
$L[\langle\bar A_0\rangle]$ quickly approaches unity above the
transition temperature can be formulated as the more precise
statement,
\begin{align}
	\langle \bar A_0\rangle\approx 0\quad \text{if} \quad T\gtrsim 1.3\, T_c\,.
\end{align}
As a consequence, we can expect quantitative effects due to the
non-trivial background only at temperatures $T\lesssim 1.3\, T_c\,$.
The most immediate effect of this non-trivial background is a shift in
the Matsubara frequencies $\omega_n \to \omega_n \pm \twopiT \, \nu_i\,$,
where $\nu_i$ are the eigenvalues of
$\bar\varphi\,$, or equivalently of $ \beta g \langle\bar A_0\rangle /(2 \pi)\,$. 
Rotating the constant field into the Cartan sub-algebra, these are given by
\begin{align}
	\label{eq:nu}
	\nu_{\text{\tiny{$SU(2)$}}} =\lbrace 0,\,\pm \varphi_3 \rbrace\,, \quad 
	\nu_{\text{\tiny{$SU(3)$}}}=\left\lbrace 0,\,0,\,\pm\varphi_3,\,
	\pm\tfrac{\varphi_3\pm\sqrt{3}\,\varphi_8}{2} \right\rbrace\,,
\end{align}
in $SU(2)$ and $SU(3)\,$, see \eg \cite{Herbst:2015ona}.  However, for
$T\lesssim 0.5\, T_c$ the effect of the shifts of the Matsubara
frequencies is suppressed by the zero temperature gapping
$\mgap$ of the gluon propagator
$2 \pi \nu_i T/\mgap\ll 1\,$. Therefore, we expect sizeable
effects due to the non-trivial background only in the regime
\begin{align}
	\label{eq:0513}
	T \in ( 0.5 \,T_c\,,\, 1.3\, T_c)\,,
\end{align}
and in particular in the electric gluon propagator.

\subsection{Debye screening mass}
\label{sec:Setup:DebyeMass}

Gluons are screened at high temperatures by the standard thermal Debye
mass.  However, also in the confined phase, they posses a finite
screening mass.  Our non-perturbative results allow to compute a
screening mass also below the critical temperature. We extract it from
the zero mode of the electric gluon propagator,
$\GE(p)=\langle A_0(p) A_0(-p))\rangle\,$, whose computation is
detailed below in \Sec{sec:Method}.  To this end, we Fourier transform
the propagator,
\begin{align}
	\label{eq:SpatialProp}
	\GE(x) = \int_{-\infty}^{\infty} \frac{\dd p}{2\pi} \; \GE(p) \; e^{\imag \, p \, x}\,.
\end{align}
At high temperatures, the screening mass can then be extracted from the
exponential decay at large distances,
\begin{align}
	\label{eq:DecayT}
	\lim_{x\to \infty} \; \GE(x) = c_e \, \exp\left(-m_s \, x\right)\,.
\end{align}
The screening mass $m_s$ obtained with \eq{eq:DecayT} is shown in
\Fig{fig:DebyeMass}.  The large distance behaviour of
$\GE(x)$ and the fits by \eq{eq:DecayT} are provided in
\App{app:Screening}.  The left panel shows
that the screening mass is finite across the phase transition and
possesses a minimum at some finite temperature. Perturbatively, the
Debye mass is given to leading order by
\begin{align}
	\label{eq:DebyeMass1}
	m^0_D=\sqrt{\frac{N}{3}} g_T T + \mathcal{O}(g_T^2 T)\,.
\end{align}
A prescription for taking higher-order effects into account has been proposed by \cite{Arnold:1995bh}
\begin{align}
	\label{eq:DebyeMass2}
  m_D=m_D^0+\left(c_D+\frac{N}{4\pi}  
  \ln\left(\frac{m_D^0}{g_T^2 T}\right)\right) g_T^2 T + \mathcal{O}(g_T^3 T)\,.
\end{align}
In order to compare our screening mass to the expressions \eq{eq:DebyeMass1} and \eq{eq:DebyeMass2}, we have to
determine $g_T$ and $c_D\,$. We use
\begin{align}
	\label{eq:Debye_pert_coupling}
	g_T=\sqrt{4\pi\,\alphaAAA^\text{E}(T,\,p=c_p\; \twopiT)}\,,
\end{align}
and fit $c_D$ at large temperatures to our result since it is not
computable within perturbation theory.  The running coupling of the
(electric) three-gluon vertex $\alphaAAA^\text{E}$ is introduced below
in \eq{eq:RunningCouplings}.

\begin{figure}[b]
	\includegraphics[width=0.55\textwidth]{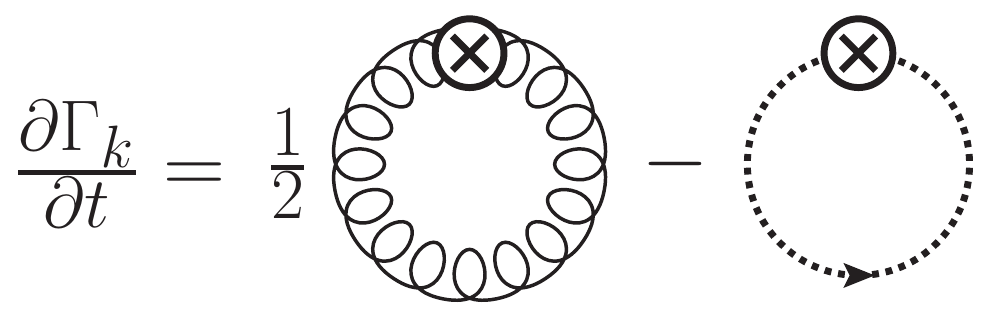}
	\caption{Graphical representation of the Wetterich equation. Wiggly (dotted) lines represent
		the dressed gluon (ghost) propagators. The cross in the circle denotes the regulator
		insertion $\partial_t R_k$ of the corresponding field type.}
	\label{fig:Wetterich}
\end{figure}

As shown in \Fig{fig:DebyeMass:HighT}, the Arnold-Yaffe Debye
mass agrees almost perfectly with our non-perturbative result down to
$T\approx\SI{0.6}{\GeV}$.  In contrast, the leading-order
Debye mass deviates instantly from our result.  By default, we set
$c_p=1$ in \eq{eq:Debye_pert_coupling} because this is the scale that
is expected to contribute most.  This yields for the non-perturbative
constant $c_D=0.100(3)\,$.  To substantiate the choice $c_p=1\,$, we
also leave it as a free fit parameter and find $c_p=0.88(64)\,$, where
the non-perturbative constant $c_D=0.105(30)$ changes only within the
fit uncertainties. The effect on the resulting Debye mass is
negligible, see \Fig{fig:DebyeMass:HighT}.
The excellent agreement at very high temperatures
provides a non-trivial check of the calculations.
Further physical consequences are discussed in \Sec{sec:Discussion:DebyeMass}.

\section{Method}
\label{sec:Method}

In this section, we discuss the flow equations, the implications
of the regulator term at non-vanishing RG scale, and we
provide details on the numerical implementation.

\begin{figure}
	\includegraphics[width=0.98\textwidth]{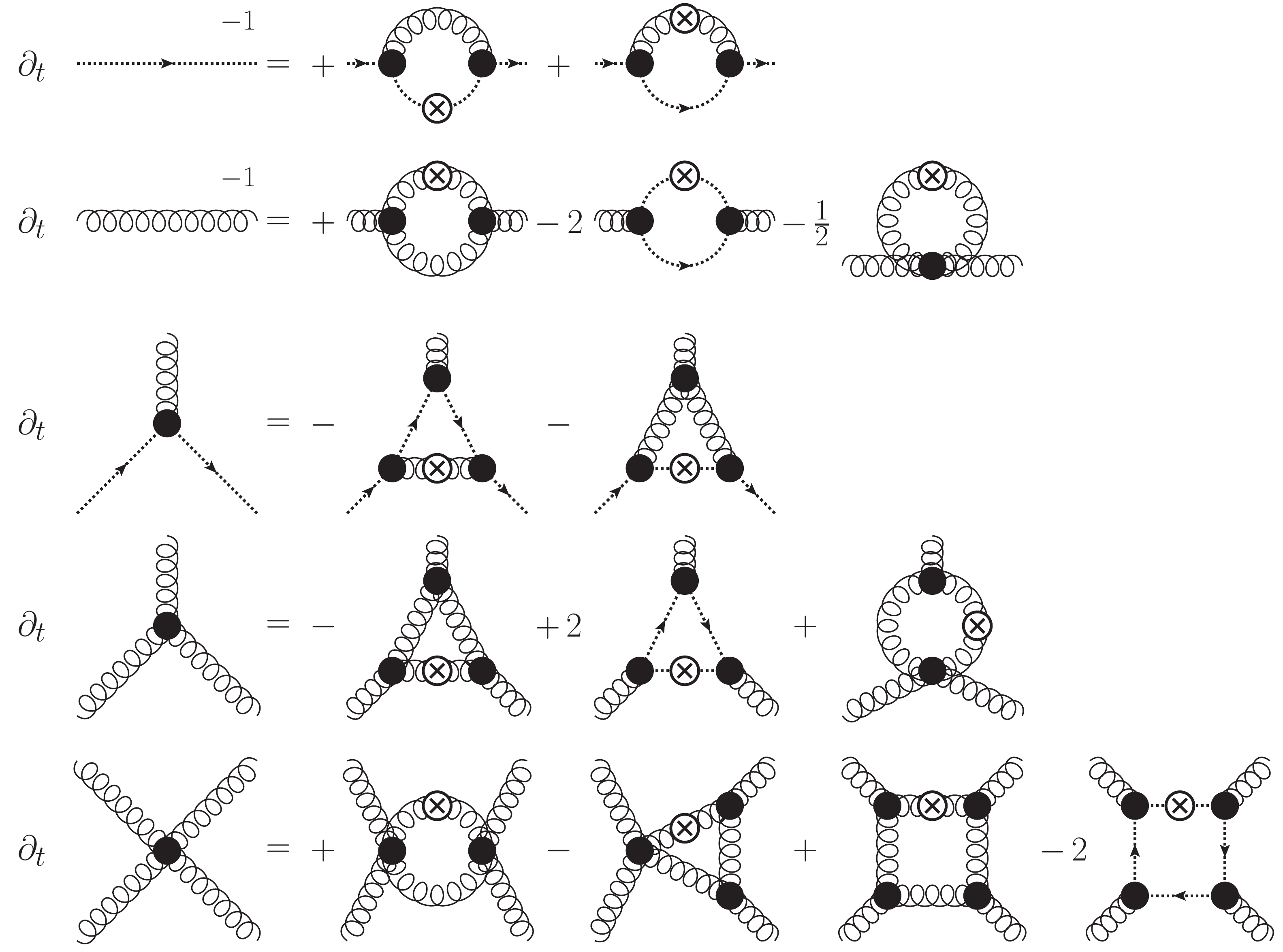}
	\caption{
		Truncated equations of all computed $n$-point 
		functions. Permutations of external legs and regulator insertions 
		are omitted in the vertex equations. 
		The gluon lines represent the sum over electric and 
			the magnetic propagators, \cf \eq{eq:propagators}.
			They are connected to the corresponding magnetic and electric 
			  components of the vertices, see also \Fig{fig:Truncation}.
		\myhfill
	}
	\label{fig:DiagrammaticFlowEquations}
\end{figure}

\begin{figure*}
	\ffigbox{
		\begin{subfloatrow}
			\ffigbox{\includegraphics{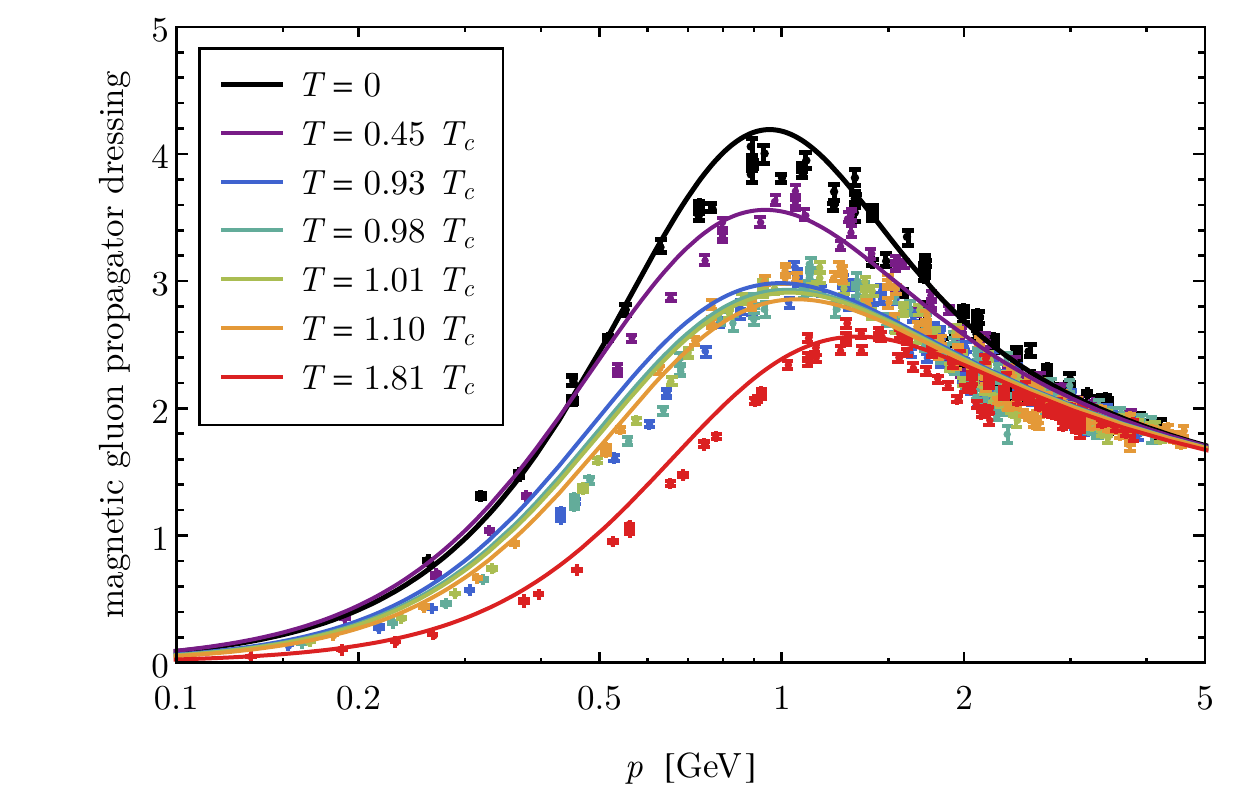}}{
				\caption{Comparison with $SU(2)$ results \cite{Maas:2011ez,Maas:PC}.}
				\label{fig:MagneticGluonProp:SU2}
			}
			\ffigbox{\includegraphics{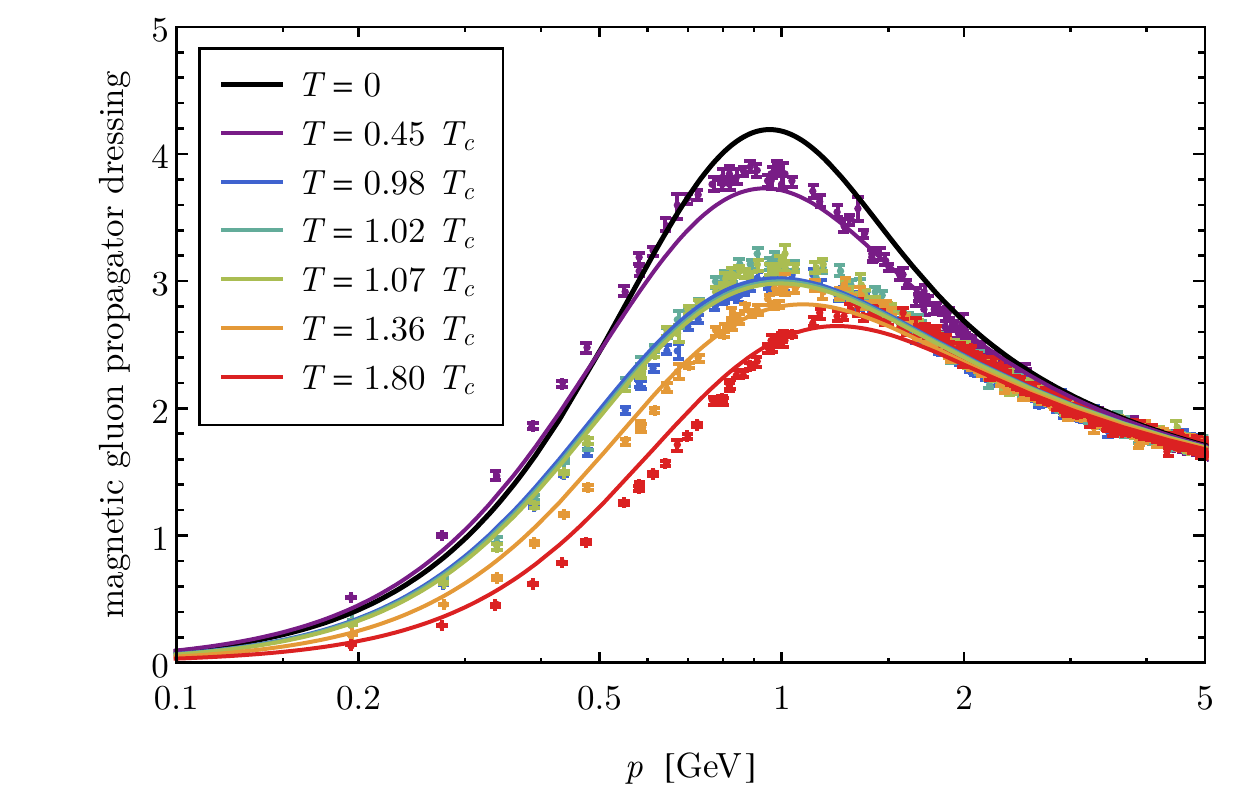}}{
				\caption{Comparison with $SU(3)$ results \cite{Silva:2013maa}.}
				\label{fig:MagneticGluonProp:SU3}
			}
		\end{subfloatrow}
	}{
		\caption{Magnetic gluon propagator dressing, \eq{eq:propagators}.}
		\label{fig:MagneticGluonProp}
	}
\end{figure*}

\subsection{FRG flows}
\label{sec:Method:FRG}

The functional renormalisation group in Wetterich's formulation 
\cite{Wetterich:1992yh} allows to integrate 
momentum-shell contributions to the effective action in the Wilsonian spirit.
To this end, a scale- and momentum-dependent mass term is added to the classical action, 
\begin{align}
	\label{eq:dSk}
	\Delta S_k=\int_x\,
	\frac{1}{2}  A_\mu^a\, R_{k,\mu\nu}^{ab} \, A_\nu^b  
		+ \int_x\,\bar c^a\, R^{ab}_k\, c^b\,.
\end{align}
The regulator term suppresses quantum as well as thermal fluctuations at momenta 
below the RG scale $k\,$. Taking the derivative with respect to the scale $k$ leads to 
the Wetterich equation for the generalised effective action $\Gamma_k\,$. This flow equation
interpolates between the classical $\Gamma_{k\rightarrow\Lambda\rightarrow\infty}=S$ 
and the $1$PI effective action $\Gamma_{k\rightarrow0}=\Gamma\,$.
For the pure gauge theory at finite temperature, the Wetterich
equation reads
\begin{align}
	\label{eq:flow}
	\partial_t \Gamma_k[\Phi] = 
		 \sumint_q \, \frac{1}{2} \ G_{k,\mu\nu}^{ab}[\Phi]\,\partial_t R_{k,\nu\mu}^{ba} 
		-\sumint_q \, G^{ab}_k[\Phi]\, \partial_t R^{ba}_k\,,
\end{align}
where $t=\ln(k/\Lambda)$ denotes the RG time and
\begin{align}
	\label{eq:Gn}
	G_k[\Phi]=\frac{1}{\Gamma^{(2)}[\Phi]+R_k}\,.
\end{align}
Using the Matsubara formalism, the momentum integral in \eq{eq:flow} is given by
\begin{align}
	\label{eq:Matsubara}
	\sumint_q = \int \frac{\dd^3 q}{(2\pi)^3} \; T \sum_{n}\,,
\end{align}
where $q_0=\twopiT\,n\equiv \omega_n\,$. A graphical representation of
the Wetterich equation for the effective action is shown in
\Fig{fig:Wetterich}.
The truncated flow equations for the correlation functions that
are obtained by taking functional derivatives of \eq{eq:flow}
are displayed in \Fig{fig:DiagrammaticFlowEquations}.

Instead of the flat regulator \cite{Litim:2000ci} 
used in the vacuum computation \cite{Cyrol:2016tym}, we use an exponential regulator. 
As demonstrated in \App{app:RegulatorIndependence}, the results for the correlation 
functions do not depend on this choice within our error bars. 
However, analytic regulators such as the
exponential regulator are better suited for numerical calculations of
thermodynamic quantities since they carry the thermal exponential decay with the cutoff scale 
$\sim e^{-c k/T}$ in the flow \cite{Fister:2011uw,Fister:Diss},
see \cite{Fister:2015eca} for a detailed study.

To reduce the numerical effort of the finite-temperature calculation, we exploit the 
degeneracy of the dressings for $k \gg \twopiT\,$. 
We integrate the finite-temperature flow starting
from the non-trivial zero-temperature effective action at 
\begin{align}
	\label{eq:init_scale_main}
	\Lambda_T&=\max\left(\lambda \, \twopiT,\, \LambdaTmin\right)\,,
\end{align}
with $\lambda=4$ and $\LambdaTmin=\SI{1}{\GeV}$, see \App{app:StartScale} for details.

\subsection{Renormalisation and mSTIs}
\label{sec:Method:Renormalisation}

In the presence of a regulator, the BRST-symmetry leads to modified
Slavnov-Taylor identities (mSTIs) for non-vanishing RG scales, $k>0$
\cite{Ellwanger:1994iz,Ellwanger:1995qf,D'Attanasio:1996jd,Litim:1998qi,Igarashi:2001mf,Pawlowski:2005xe,Igarashi:2016gcf,Cyrol:2016tym}.
The additional terms are generated by the BRST-variation of the
regulator term and have a one loop form. They are similar in form and
structure to the flow equation itself. The latter encodes the breaking
or flow of scale invariance while the former encode the breaking or
flow of BRST symmetry. The resulting mSTIs reduce to the standard STIs
in the limit of vanishing RG scale, $k\rightarrow 0\,$, similar to the
removal of the explicit breaking of scale invariance due to the
regulator. Therefore, in both cases the underlying symmetry is restored
in the limit of vanishing RG scale $k\rightarrow 0\,$. 
We emphasise that any regularisation scheme in momentum space leads to 
such a modification of BRST symmetry in terms of modified STIs.
This is also well known from perturbation theory, where a
cutoff regularisation, amongst other modifications, requires a gluon
mass counter term in order to guarantee gauge invariance. Modified STIs
are also present within other functional methods
such as non-perturbative DSE and $n$PI approaches
that rely on numerical momentum integrations.

\begin{figure*}
	\ffigbox{
		\begin{subfloatrow}
			\ffigbox{\includegraphics{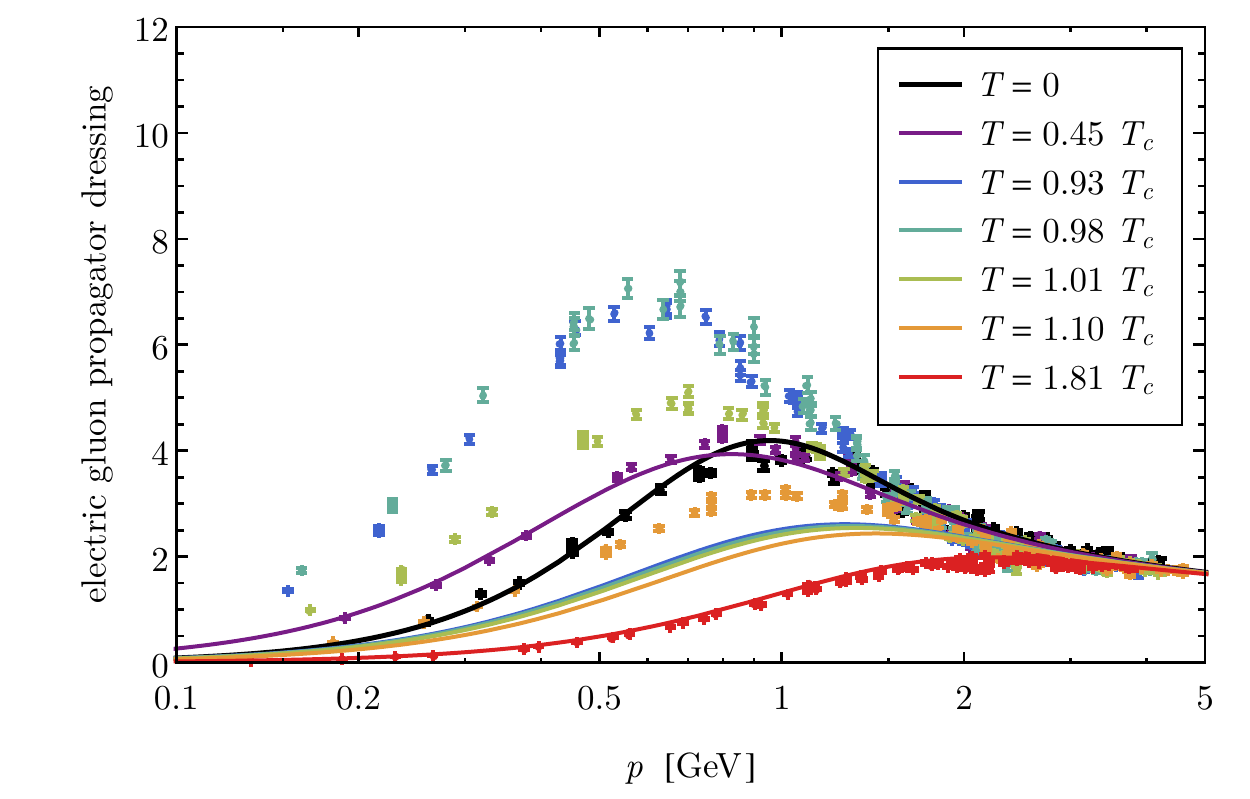}}{
				\caption{Comparison with $SU(2)$ results \cite{Maas:2011ez,Maas:PC}.}
				\label{fig:ElectricGluonProp:SU2}
			}
			\ffigbox{\includegraphics{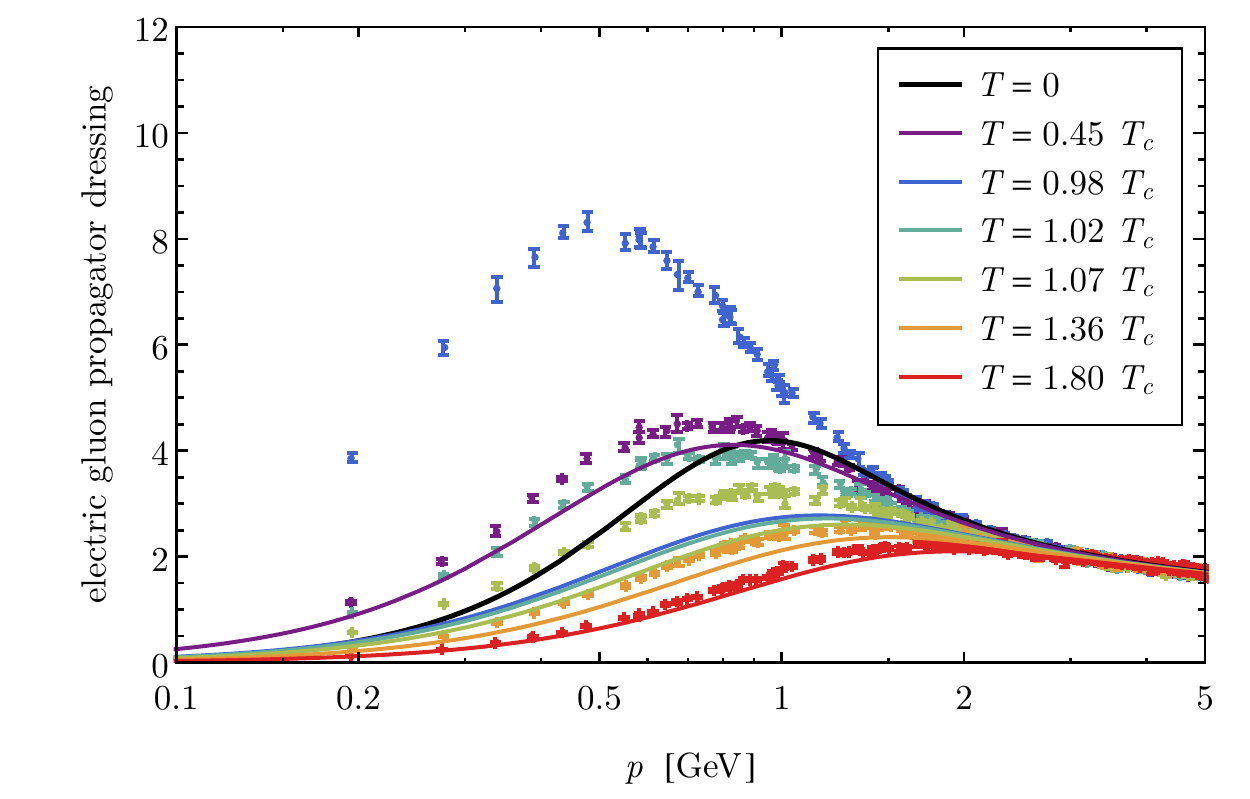}}{
				\caption{Comparison with $SU(3)$ results \cite{Silva:2013maa}.}
				\label{fig:ElectricGluonProp:SU3}
			}
		\end{subfloatrow}
	}{
		\caption{Electric gluon propagator dressing, \eq{eq:propagators}.}
		\label{fig:ElectricGluonProp}
	}
\end{figure*}

To take the modification of the STIs of the vertices into account, we choose constant
vertex dressings $\lAcc\,$, $\lAAA$ and $\lAAAA$ at the cutoff scale, $k=\Lambda\,$,
such that the STIs for the running couplings,
\begin{align}
	\label{eq:alpha}
	\alphaAcc(\mu)=\alphaAAA(\mu)=\alphaAAAA(\mu)\equiv\alpha(\mu)\,,
\end{align}
are fulfilled at $\mu=\SI{20}{\GeV}$, $k=0\,$.
Here, the running couplings in \eq{eq:alpha} are obtained from the classical vertices,
\begin{align}
	\label{eq:RunningCouplings}
	\alphaAcc(\psym) &= \frac{\left(\lAcc(\psym)\right)^2}{Z_A(\psym)\,\Zc^2(\psym)}\,,\eqnewline
	\alphaAAA(\psym) &= \frac{\left(\lAAA(\psym)\right)^2}{\ZA^3(\psym)}\,,\eqnewline
	\alphaAAAA(\psym) &= \frac{\lAAAA(\psym)}{\ZA^2(\psym)}\,,
\end{align}
with the symmetric momentum configuration $\psym\,$.

The mSTI of the gluon propagator implies a non-vanishing longitudinal
gluon mass term at the cutoff scale \cite{Ellwanger:1994iz}. In the
perturbative regime, it can be shown that the transverse mass agrees
with the longitudinal one, for details see
\cite{Cyrol:2016tym}. However, while the longitudinal mass parameter
vanishes at $k=0\,$, the transverse mass term encodes the gapping of
the transverse gluon propagator at $k=0\,$. At the initial UV cutoff scale
$k=\Lambda$ the gluon mass parameter is uniquely determined by the
mSTI and cannot be chosen freely. Its precise determination is at the
root of confinement, which is encoded in the transverse mass gap at
vanishing cutoff scale. Since the mass parameter is proportional to
the cutoff, $m_\Lambda^2\propto\alpha(\Lambda)\,\Lambda^2\,$,
quadratic precision is required in its determination from the mSTI.
The solution of this quadratic fine tuning problem requires both, a
BRST-consistent quantitative level of the approximation, as well as 
sufficient numerical precision. Consequently, in truncated systems of
flow equations, its computation from the mSTI at the required
precision level is extremely challenging. The above brief discussion
is detailed in \cite{Cyrol:2016tym}. Note that these statements
hold also for other functional methods such as DSE and $n$PI
approaches.

In the present work we utilise that it is possible to uniquely
determine the gluon mass parameter by demanding a solution of the
scaling type, for details see again \cite{Cyrol:2016tym}. 
We exploit that this also holds at finite temperature.
Requiring scaling in the magnetic sector provides us with a unique
value for the gluon mass parameter at each temperature. This procedure
resolves the necessity of a BRST-consistent level of the approximation,
but still requires quadratic precision in the fine-tuning. Further
details are provided in \App{app:UVmass}.

\begin{figure*}
	\ffigbox{
		\begin{subfloatrow}
			\ffigbox{\includegraphics{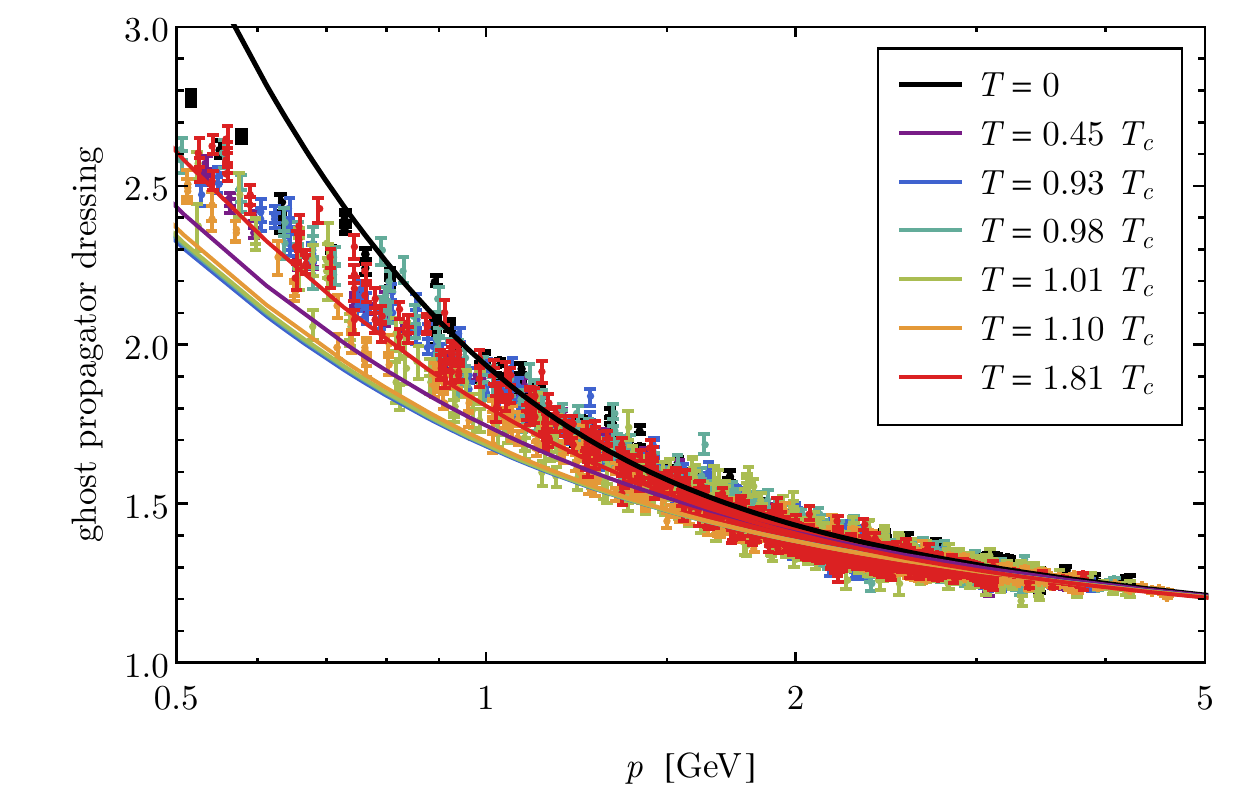}}{
				\caption{Ghost propagator dressing $1/\Zc(p)\,$.}
				\label{fig:GhostDressings:Propagator}
			}
			\ffigbox{\includegraphics{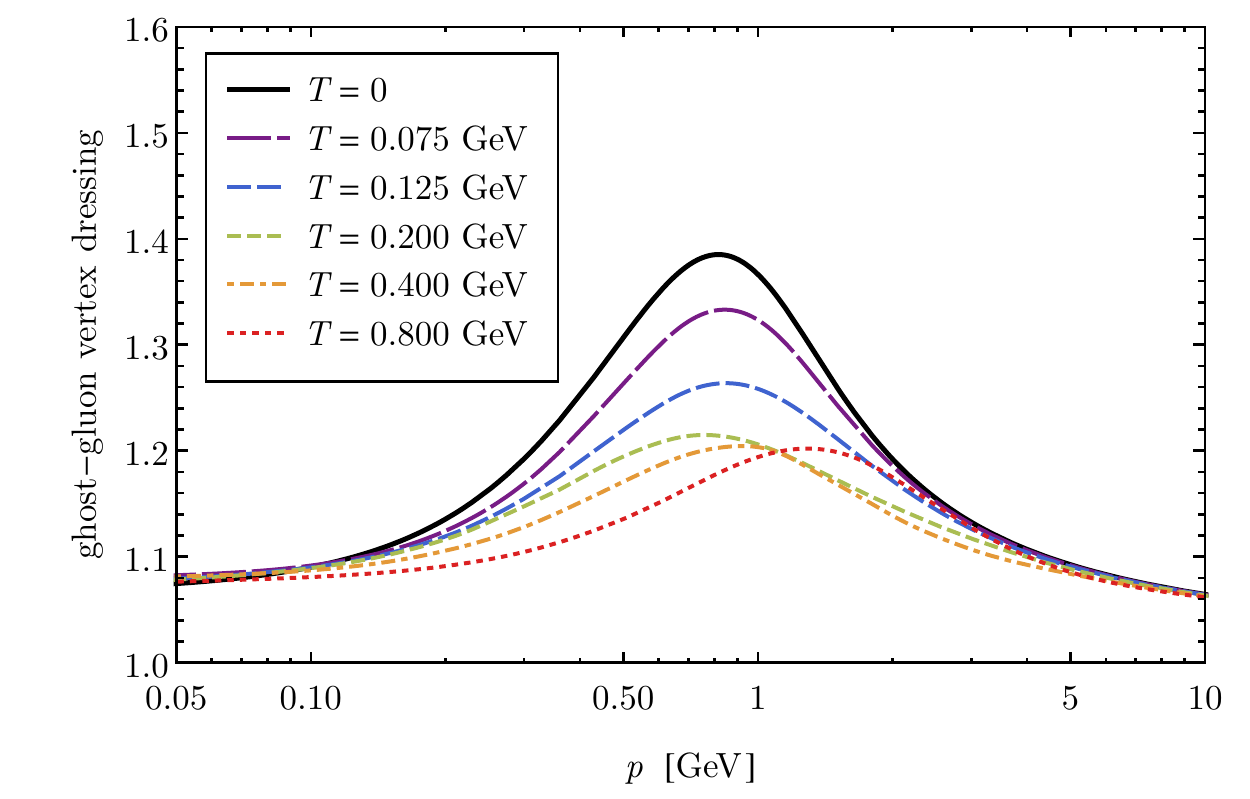}}{
				\caption{Magnetic ghost-gluon vertex dressing $\lAccM(p)\,$.}
				\label{fig:GhostDressings:Vertex}
			}
		\end{subfloatrow}
	}{
	\caption{
		Ghost propagator dressing, \eq{eq:propagators}, compared to
		$SU(2)$ lattice results 
		\cite{Maas:2011ez,Maas:PC}
		and ghost-gluon vertex, \eq{eq:GhostGluonVertex}.
	}
	\label{fig:GhostDressings}
}
\end{figure*}

\subsection{Numerical implementation}
\label{sec:Setup:NumericalCalculation}

To solve the system of coupled flow equations, 
we use the tools established by the fQCD collaboration~\cite{fQCD:2016-10}.
The tensorial flow equations are derived with DoFun~\cite{Huber:2011qr}.
Subsequently, the projected equations are traced with FormTracer 
\cite{Cyrol:2016zqb},
a Mathematica package that uses FORM 
\cite{Vermaseren:2000nd,Kuipers:2012rf,Kuipers:2013pba,Ruijl:2017dtg}
and has native support 
for finite-temperature applications. The output is exported as optimised \CC 
code, which is then used within the computational framework of the fQCD
collaboration. The latter uses the adaptive ordinary 
differential equation solver from the Boost library \cite{odeboost} and 
the adaptive multidimensional integration routine from \cite{cubature},
which implements \cite{berntsen1991adaptive,genz1980remarks}.

In the derivation of the equations, tracing the four-gluon vertex
equation, and in particular the gluon box diagrams, is the most
challenging part. To this end, we generate FORM files with FormTracer
\cite{Cyrol:2016zqb} for each of the twelve permutations of the box
diagrams.  Executing one of these with FORM can take up to eight core
days and intermediate expressions reach more than \SI{1}{\tera\byte}
in size.  Since the resulting expressions are still very large, we sum
all permutations, factorise all dressing functions and then use the
simultaneous optimisation feature of FORM's optimisation routine
\cite{Kuipers:2013pba} in combination with a parallelised version of
FORM \cite{Tentyukov:2007mu} to optimise the result. Concerning the
numerical computation, integrating the flow once takes roughly one day
on an ordinary quad-core desktop computer. This has to be done
multiple times for each temperature due to the gluon mass parameter
determination.

\section{Results}
\label{sec:Results}

\begin{figure*}
	\ffigbox{
		\begin{subfloatrow}
			\ffigbox{\includegraphics{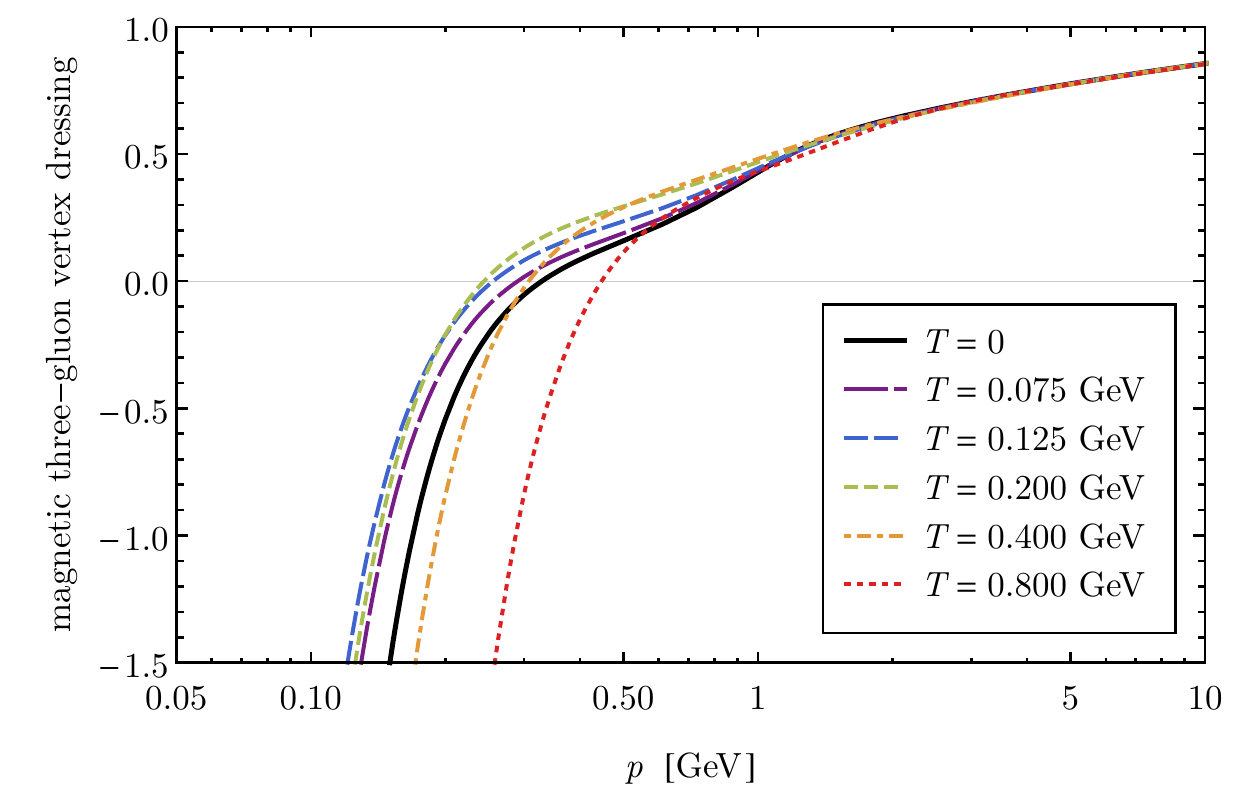}}{
				\caption{Magnetic three-gluon vertex dressing $\lAAAM(p)\,$.}
				\label{fig:GluonicVertices:MagneticAAA}
			} 
			\ffigbox{\includegraphics{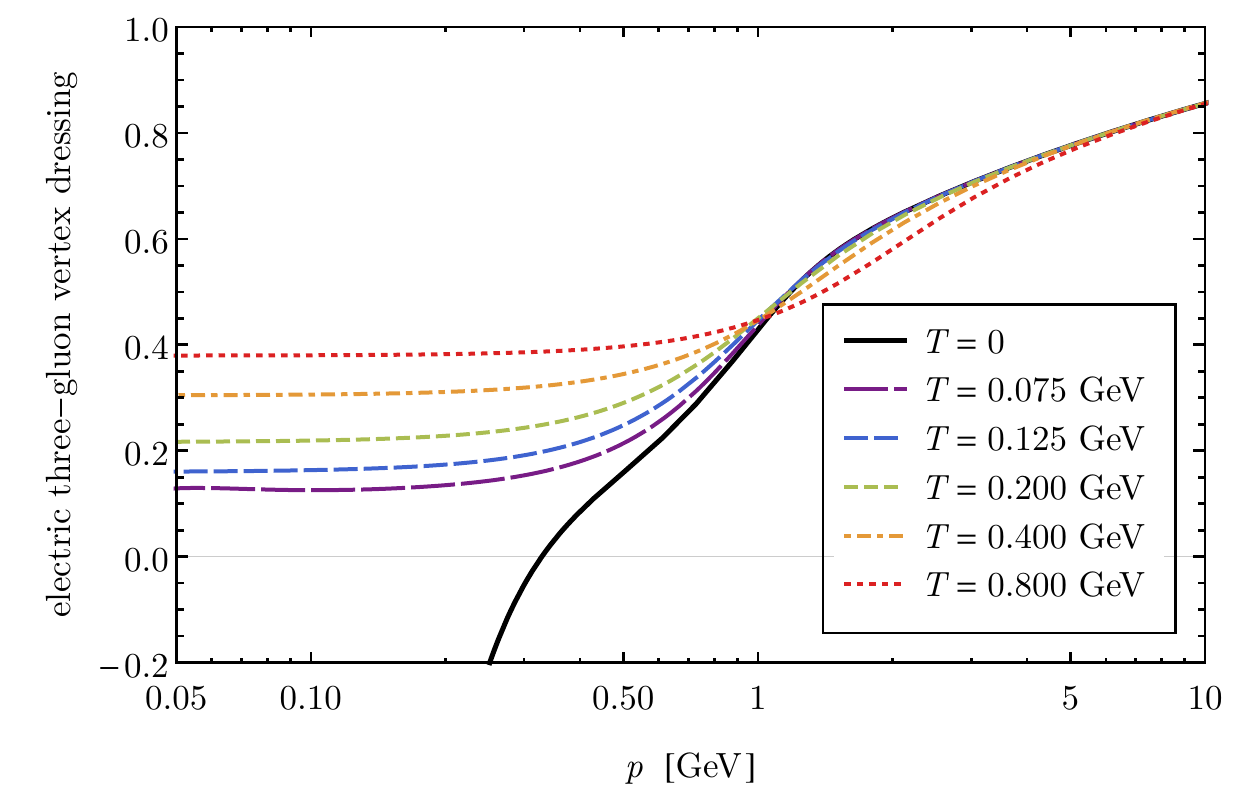}}{
				\caption{Electric three-gluon vertex dressing $\lAAAE(p)\,$.}
				\label{fig:GluonicVertices:ElectricAAA}
			}
		\end{subfloatrow}
	}{
	\caption{Temperature dependence of the three-gluon vertex dressing, \eq{eq:ThreeGluonVertex}.}
	\label{fig:ThreeGluonVertex}
}
\end{figure*}

\begin{figure*}
	\ffigbox{
		\begin{subfloatrow}
			\ffigbox{\includegraphics{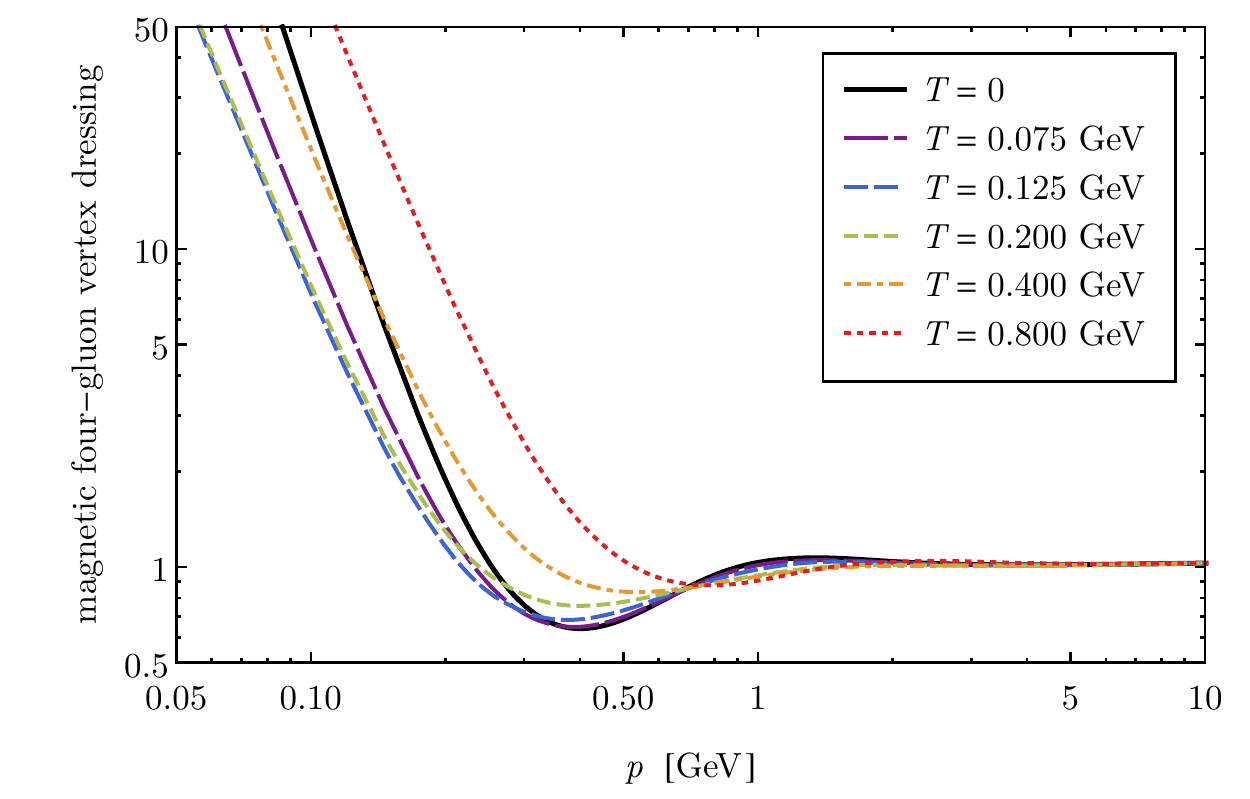}}{
				\caption{Magnetic four-gluon vertex dressing $\lAAAAM(p)\,$.}
				\label{fig:GluonicVertices:MagneticAAAA}
			} 
			\ffigbox{\includegraphics{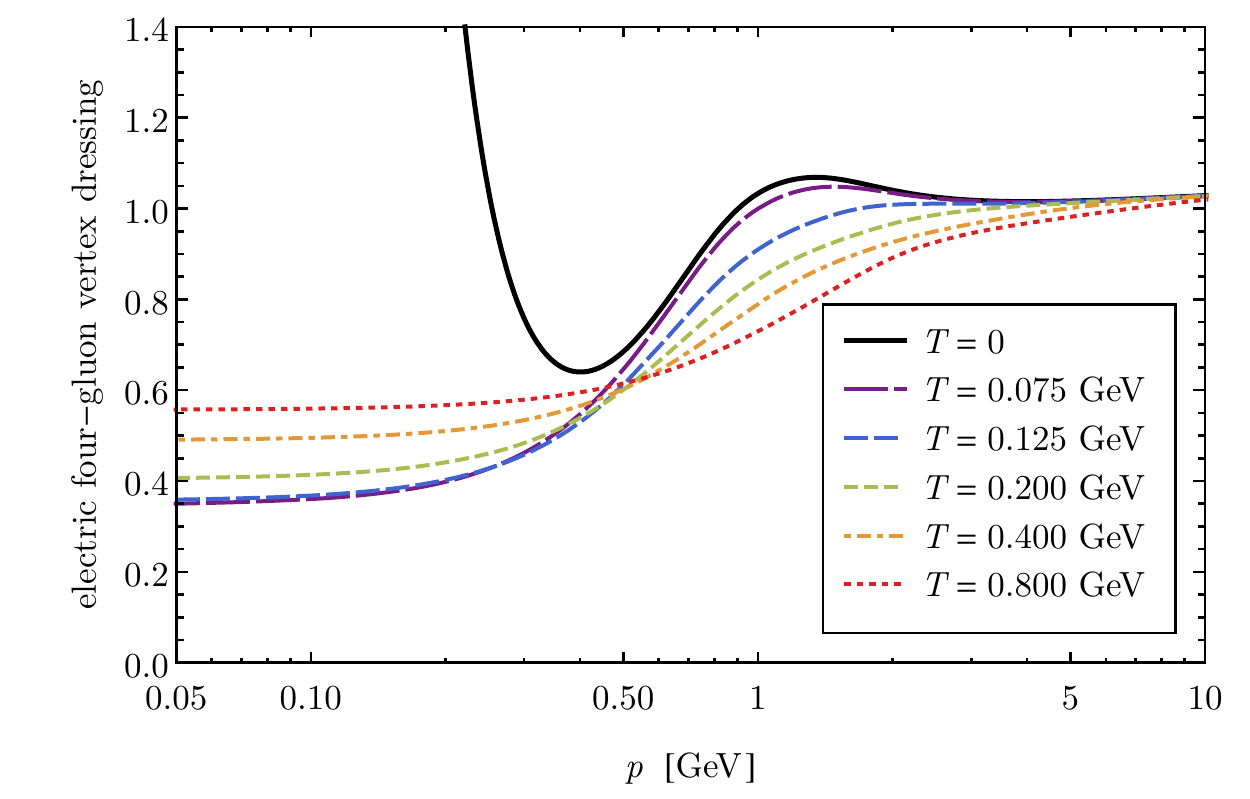}}{
				\caption{Electric four-gluon vertex dressing $\lAAAAE(p)\,$.}
				\label{fig:GluonicVertices:ElectricAAAA}
			}
		\end{subfloatrow}
	}{
	\caption{Temperature dependence of the four-gluon vertex dressing, \eq{eq:FourGluonVertex}.}
	\label{fig:FourGluonVertex}
}
\end{figure*}

The main results are displayed in
\Figrange{fig:MagneticGluonProp}{fig:ZeroCrossing}.  We show
results for the magnetic and electric dressing functions of
propagators and vertices for various temperatures. For all
correlators we find that the magnetic and electric dressings coincide
for momenta $p\gg \twopiT\,$, and become degenerate with the vacuum
dressings. This is required by the recovery of $O(4)$ invariance. 
The convergence towards the vacuum dressings for small temperatures
is explicitly checked in \App{app:VacuumLimit}. This apparently obvious
property is actually non-trivial within frequency and momentum-dependent
non-perturbative truncations.

We compare our gluon propagators to
$SU(2)$~\cite{Maas:2011ez,Maas:PC} and $SU(3)$~\cite{Silva:2013maa}
lattice results in
\Figs{fig:MagneticGluonProp}{fig:ElectricGluonProp}. 
This comparison requires the setting of a relative scale as well as
renormalisation, detailed in \App{app:ScaleSettingAndRenormalisation}. 
As a consequence, a potential relative offset of functional and lattice 
results has to be considered in addition to the systematic errors of the
truncation, when juxtaposing the results from the different calculations.
The comparison with
$SU(2)$ as well as $SU(3)$ lattice data is legitimate because
the truncation used in this work yields only a trivial dependence on the gauge group.
This is the case because the colour traces can be taken 
without specifying the gauge group \cite{vanRitbergen:1998pn,Cyrol:2016zqb}, 
and the only group constant appearing in the equations is the
quadratic Casimir operator of the adjoint representation $C_A\,$. 
Furthermore, $C_A$ occurs only in combination with the coupling
at the renormalisation point
$\alpha(\mu) \cdot C_A \equiv \tilde{\alpha}(\mu)\,$. Thus, it can be
absorbed into a redefinition of the running coupling, or, equivalently,
the scale of the theory. Therefore, the propagators are identical for all groups,
and the different couplings can be obtained by a trivial rescaling
with $C_A\,$. We emphasize that this is not a mere artefact of the
approximation. Perturbatively, the beta function of the pure gauge
theory has a trivial group dependence up to three loops, see \eg
\cite{Herzog:2017ohr} for a recent discussion.
See \Sec{sec:Discussion} for a discussion.

At low momenta, the electric and magnetic propagators show a
qualitatively different behaviour. While the magnetic gluon
propagator decreases almost monotonously with increasing
temperature, the electric propagator increases at small
temperatures.  At high temperatures, where the growth of thermal
contributions to the mass becomes dominant, also the electric gluon
propagator decreases, see also \Sec{sec:Setup:DebyeMass} and in particular \Fig{fig:DebyeMass}.
For the magnetic gluon propagator we find agreement with the lattice results 
on the \SI{10}{\percent} accuracy level we expect from the truncation of the vertices. 
Furthermore, we see that the deviation takes its maximum for temperatures about the
phase transition temperature, where we 
expect large-scale dynamical fluctuations to be most relevant. 
On the one hand, our truncation is tested maximally in this regime,
and on the other hand discretisation and finite volume effects in the lattice calculation
are strongest there. In contradistinction to
the very satisfactory situation for the magnetic propagator, we
observe a significant deviation about the phase transition
temperature $T_c$ for the electric gluon propagator. However, the
agreement is very good for small and large temperatures. 
As discussed in great detail in \Secs{sec:Setup:ExpansionPoint}{sec:Discussion},
the deviation about $T_c$ can be explained by the missing non-trivial
$\langle A_0\rangle$-background in the present calculation.

\begin{table}[b]
	\def\arraystretch{1.2}
	\begin{tabular}{l||l|l|l}
		& \makecell{simple vertices}
		& \makecell{sym. mom.\\ approx.}
		& \makecell{full mom.\\dependence}\\
		\hline
		$d=4$ & $\,0.5953\,$\cite{Zwanziger:2001kw,Lerche:2002ep,Pawlowski:2003hq} & $\,0.567(3)$ & $\,0.576(5)\,$\cite{Cyrol:2016tym} \\
		$d=3$ & $\,0.3976\,$\cite{Zwanziger:2001kw} & $\,0.321(1)\,$\cite{Corell:Prep} & \makecell{$\times$}\\
		$d=4,\,T>0$ & \makecell{$\times$}  & $\,0.323(3)$ & \makecell{$\times$} 
	\end{tabular}
	\caption{Scaling exponents. See \Sec{sec:Setup:VertexExpansion} for the 
		definition of the symmetric momentum approximation and 
		\App{app:RegulatorIndependence} for a comparison.\myhfill
	}
	\label{tab:ScalingExponents}
\end{table}

The ghost propagator agrees qualitatively, but 
deviates quantitatively, from the lattice results, as shown in  \Fig{fig:GhostDressings:Propagator}.
We discuss this point further in \Sec{sec:Discussion}.
The ghost-gluon vertex is plotted in \Fig{fig:GhostDressings:Vertex}.
Interestingly, it is weaker around the phase 
transition temperature than in the vacuum. At high temperatures it shows a 
broader and less pronounced bump than at zero temperature.

The gluonic vertex dressing functions are shown in \Figs{fig:ThreeGluonVertex}{fig:FourGluonVertex}.
The magnetic dressings of both vertices show scaling in the infrared.
Contrarily, the corresponding electric components decouple at $p\approx \twopiT$ and become 
constant in the infrared.
We show the position of the zero crossing of the magnetic three-gluon vertex dressing function as a function of temperature in 
\Fig{fig:ZeroCrossing}. At small temperatures the zero 
crossing moves towards lower momenta as the temperature is increased,
since the three-gluon vertex is stronger at small and intermediate momenta
for small temperatures, \cf \Fig{fig:ThreeGluonVertex}.
At high temperatures, the magnetic zero crossing rises linearly with the temperature.
In contrast, the zero crossing of the electric three-gluon vertex dressing function
disappears at $T\approx\SI{40}{\MeV}$.
Similarly, the electric dressing of the four-gluon vertex undergoes a 
drastic change from zero to small temperatures, where scaling is 
lost, and goes on to increase with growing temperature.

At low momenta $p\ll \twopiT$ the dimension of the theory is effectively reduced
and the magnetic dressings behave as they do in three dimensions. 
In the case of the scaling solution, all magnetic dressing functions scale 
with a power-law~\cite{Fischer:2006vf,Huber:2007kc,Fischer:2009tn},
\begin{align}
	\label{eq:GeneralScaling}
	\lim_{p\rightarrow 0} \; \lambda^{(2n,m)}(p)\propto 
		\left(p^2\right)^{(n-m)\kappa+(1-n)\left(\frac{d}{2}-2\right)}\,,
\end{align}
where $2n$ and $m$ is the number of ghost and gluon legs,
respectively. Due to dimensional reduction, the temperature-independent scaling coefficient
$\kappa$ is determined by three-dimensional Yang-Mills theory. Fitting the magnetic gluon propagators to
\eq{eq:GeneralScaling} with $d=3$ at $p\ll \twopiT\,$, we find
$\kappa=0.323(3)\,$. This agrees with the scaling exponent
$\kappa=0.321(1)$ of the three-dimensional vacuum theory
\cite{Corell:Prep}.  We summarise the different scaling coefficients
in \Tab{tab:ScalingExponents}.

\section{Discussion}
\label{sec:Discussion}

In the previous section we have presented non-perturbative results
obtained with the most comprehensive truncation within
functional methods to date.
The agreement of the magnetic propagator and the
electric propagator for high temperatures is of the order of
\SI{10}{\percent} in the momentum regime relevant for hadronic observables.  These small deviations can be attributed to lattice
artefacts, the relative scale setting uncertainty, and the systematic
error within our truncation. The latter stems from incomplete
momentum dependencies of the vertices and missing non-classical
tensors, see \Apps{app:RegulatorIndependence}{app:Projections} for
estimates of their respective importance.
The electric propagator deviates from the lattice results
at temperatures about the phase transition temperature. The
explanation has already been indicated in \Sec{sec:Results} and is
discussed below. 

\subsection{Non-trivial backgrounds and their impact on electric and
  magnetic propagators}
\label{sec:Discussion:Propagators}

A potential source of the discrepancy of the electric
gluon propagator near the phase transition temperature
is an insufficient order in our approximation scheme.
However, such deviations of the electric 
gluon propagator from lattice results were already 
observed in much simpler truncations \cite{Fister:2011uw}.
Furthermore, if truncation artefacts were the main source,
we would expect larger discrepancies also in the
magnetic gluon propagator.

In contrast to this, the electric propagator, which is closely related
to the order parameter $L[\langle\bar A_0\rangle]$, is particularly sensitive to a 
non-vanishing background field \cite{Reinosa:2016iml}.
As argued in \Sec{sec:Setup:ExpansionPoint}, the non-trivial solution 
of the equation of motion,
$\bar A_0\neq 0\,$, is important in the temperature regime \eq{eq:0513},
that is
\begin{align}
	T \in ( 0.5 \,T_c\,,\, 1.3\, T_c)\,.\nonumber
\end{align}
This is exactly the temperature range where the deviations from the
lattice results, which are evaluated on the equation of motion, are
most pronounced.  We expect a considerable improvement in the electric
propagator if the correlation functions are evaluated on the
non-trivial background.  At this point, we want to emphasise that the
observed deviations do not invalidate our results for the electric
two-point correlator. It simply represents the correlation functions
at a non-minimal configuration, \cf \eq{eq:exp2}.  Furthermore, these
findings underline that Polyakov-enhanced low-energy effective models
should be constructed in $\bar A_0$-backgrounds and the effective
potential $V[\bar A_0]$ rather than Polyakov loop backgrounds and the
Polyakov loop potential $V[L]\,$: the electric propagators agree on
the \SI{10}{\percent} level above $T \gtrsim 1.3\, T_c\,$. This entails
that the relevant background for the shifts in the Matsubara
frequencies is $\langle \bar A_0\rangle\,$. 

The above analysis is also important for the discussion of the
comparison of the present results with $SU(2)$ and $SU(3)$ lattice
simulations. As discussed in detail in the last \Sec{sec:Results}, the gauge
group enters only at very high orders of the approximation in an expansion
of the effective action around vanishing background. Thus, our
results depend only trivially on the gauge group.  However, the gauge
group, and in particular the universality class, enters via the
Polyakov loop background, or, more precisely $\langle \bar A_0\rangle\,$.
 It has already been shown in
\cite{Braun:2007bx,Braun:2010cy} that the different orders of the
phase transition for $SU(2)$ and $SU(N>2)$ are encoded in the Polyakov
loop potential $V[\bar A_0]$ and the respective expectation values
$\nu$ in \eq{eq:nu}, rather than in the propagators. The Ising critical
exponents for $SU(2)$ are also extracted from critical fluctuations
encoded in the effective potential, see \cite{Marhauser:2008fz} for
Yang-Mills theory in Polyakov gauge and~\cite{Spallek:MSc} for
Landau-gauge Yang-Mills theory. In~\cite{Marhauser:2008fz,Spallek:MSc}
it also has been shown, that the critical fluctuations are the actual cause
of the higher phase transition temperature in comparison to
$SU(N>2)\,$. Thus, the gauge group dependence of the order of
the phase transition and the value of transition temperature are to leading
order caused by the effective potential, and hence by the related
expansion about the physical ground state, \ie $\langle \bar A_0\rangle$
in the current setting.

We close the discussion of the propagators with the remark that the
comparison of our results with the lattice results at small momenta
$p^2\ll \lqcd^2$ has to be taken with a grain of
salt. The lattice results are of the decoupling type, while our results
are of the scaling type. Consequently, possible
non-perturbative gauge fixing effects have to be kept in mind, see \eg
\cite{Maas:2009se,Sternbeck:2012mf,Maas:2015nva,Maas:2017csm}.  This
concerns in particular the ghost propagator, shown in
\Fig{fig:GhostDressings:Propagator}, which is more
sensitive to the treatment of the Gribov copies than the gluon
propagator \cite{Maas:2017csm}.

\subsection{Debye mass and perturbative regime}
\label{sec:Discussion:DebyeMass}

We find very good agreement of our non-perturbative Debye screening
mass with two-loop hard thermal loop perturbation theory down to
$T\approx\SI{0.6}{\GeV}$, see \Fig{fig:DebyeMass}. This remarkable
agreement down to comparably low temperatures is in line with
earlier findings, see \eg \cite{Heller:1997nqa,Datta:1999yu,Cucchieri:2001tw,Nakamura:2003pu,Arnold:1995bh,Braaten:1995ju,Braaten:1995jr,Andersen:2009tc,Andersen:2010ct}.
In general, perturbative resummation schemes have been found
to be applicable at surprisingly large couplings.
An explanation of this unexpectedly large range of
validity can be given by the structural similarity of higher order
perturbative resummation schemes and the non-perturbative resummations
performed within functional methods. This opens the door for
applications of functionally assisted analytic perturbative
computations beyond the validity bounds of perturbation theory, in
particular to the transport and kinetic realm of heavy ion collisions.

\subsection{Three-gluon vertex and its zero crossing}
\label{sec:Discussion:ZeroCrossing}

The magnetic three-gluon vertex dressing function has been studied on
the lattice \cite{Fister:2014bpa} and with a semi-perturbative
approximation of its DSE \cite{Huber:2016xbs}.  Both studies show a
significant enhancement of the magnetic dressing at low momenta
$p\approx\SI{0.2}{\GeV}$ for temperatures just below the critical
temperature.  While we also observe this effect qualitatively, see
\Fig{fig:ThreeGluonVertex}, we find a much weaker enhancement.  This
is consistent with the finding that our electric gluon propagator is
weaker than the electric lattice propagator, \cf
\Fig{fig:ElectricGluonProp}.
This electric propagator enters the triangle diagram in the
three-gluon vertex equation, which yields a positive contribution to
the dressing function \cite{Eichmann:2014xya}.  Thus, a stronger
electric propagator increases the strength of the magnetic three-gluon
vertex.

At zero temperature, the three-gluon vertex shows a zero crossing in
four as well as in three dimensions
\cite{Cucchieri:2006tf,Cucchieri:2008qm,Tissier:2011ey,Pelaez:2013cpa,%
  Aguilar:2013vaa,Blum:2014gna,Eichmann:2014xya,Huber:2016tvc,%
  Cyrol:2016tym,Athenodorou:2016oyh,Duarte:2016ieu,%
  Athenodorou:2016oyh,Boucaud:2017obn}.  Analytical studies show that
it is caused by the divergent ghost triangle diagram.  We find that
the zero crossing persists in the magnetic dressing function for all
temperatures.  This stands in line with \cite{Huber:2016xbs} but in
contrast to \cite{Fister:2014bpa}, where the lowest investigated
momenta show a positive sign at temperatures somewhat below the
critical temperature. Here, we present an analytical argument for the
persistence of the magnetic zero crossing at all temperatures. The
argument is first presented for a vanishing gluonic background and is based
on the infrared dominance of ghost loops. Finally we discuss the case of non-vanishing 
gluonic backgrounds relevant for temperatures about $T_c\,$. 

\begin{figure}
	\includegraphics[width=\textwidth]{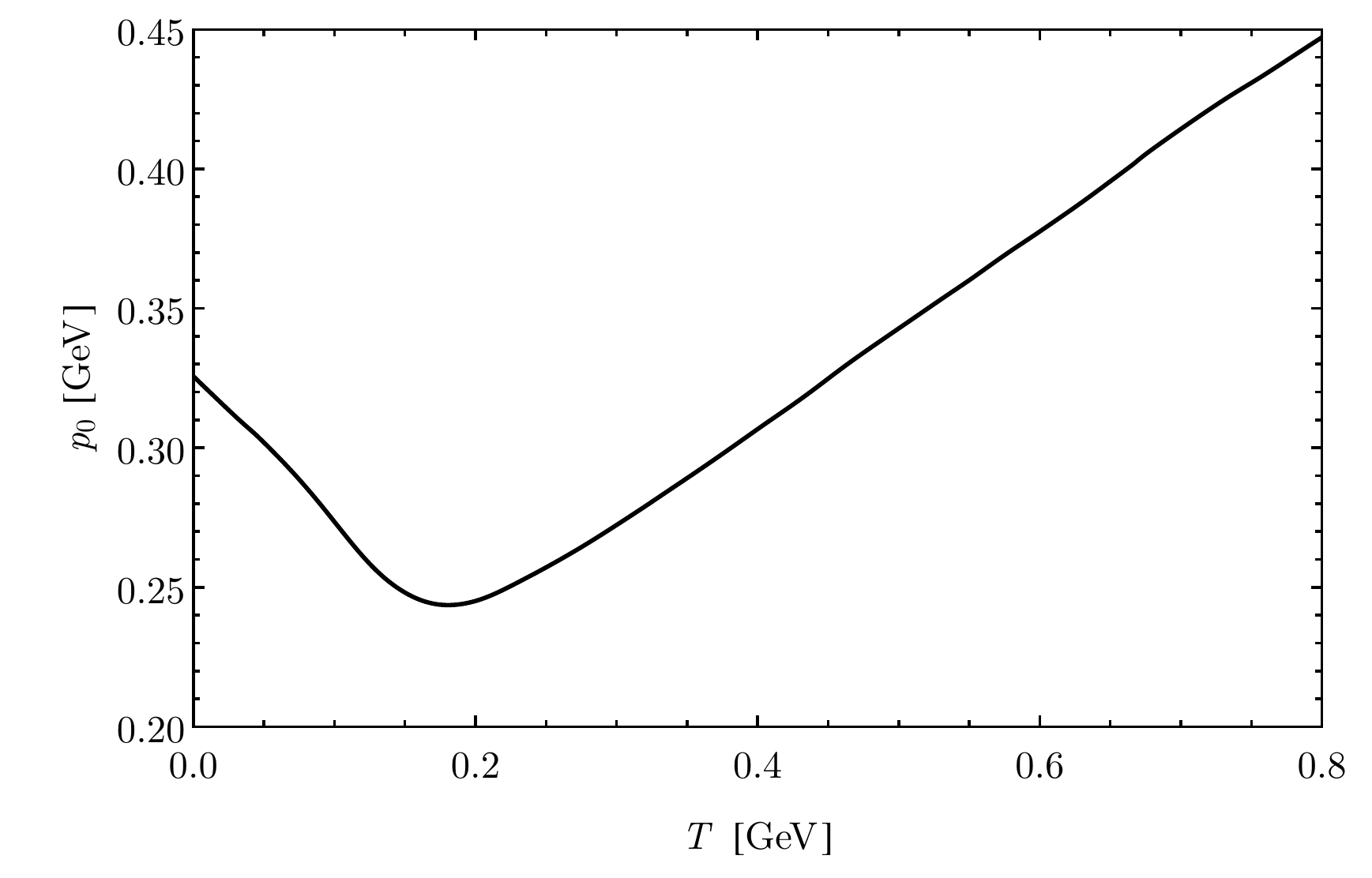}
	\caption{
		Temperature dependence of the magnetic three-gluon vertex zero crossing $\lAAAM(p_{0})=0\,$.\myhfill}
	\label{fig:ZeroCrossing}
\end{figure}

All gluonic diagrams are gapped below a certain
scale, whereas the ghost triangle effectively behaves like the
corresponding three-dimensional diagram for $p^2\ll (\twopiT)^2\,$.  Therefore, it
causes a divergence in the magnetic three-gluon vertex dressing
function at low momenta for all temperatures, and thus, the magnetic
zero crossing cannot vanish. At high temperatures, this zero crossing
moves then to higher scales, which is in line with the high
temperature limit and \cite{Fister:2014bpa}. This qualitative argument
is actually independent of the type of the solution, since the
three-dimensional ghost triangle diagram diverges with a power-law in
the case of the scaling solution and linearly
\cite{Pelaez:2013cpa,Aguilar:2013vaa,Huber:2016tvc} in the case of the
decoupling solution. We find that the zero crossing of the electric
component vanishes at a temperature of $T\approx\SI{40}{\MeV}$.  This
can be understood by observing that the zero mode of the ghost
triangle diagram, evaluated at zero external Matsubara frequencies,
contributes to the magnetic three-gluon vertex dressing, but vanishes
analytically if projected with the electric three-gluon vertex
projection operator.  Our numerical results show precisely the
expected behaviour, see
\Figs{fig:ZeroCrossing}{fig:ThreeGluonVertex}.

We extend the argument to the case of non-vanishing
backgrounds. They introduce a colour structure in the ghost propagator
and the ghost-gluon vertex. After diagonalisation, we are left with
gapped and ungapped modes in the ghost propagator, as well as
background-dependent and background-independent (colour) tensor
structures in the ghost-gluon vertex. The remaining ungapped ghost
modes couple to the latter tensor structure, which is nothing but the 
original tensor structure at vanishing background.
Therefore, the background simply leads to a weakening of
the infrared dominance by gapping some, but not all, ghost modes.
Accordingly, the zero 
crossing moves towards smaller momenta, but does not disappear, in the
presence of non-trivial backgrounds. Furthermore,
for small temperatures $T/\lqcd\to 0\,$, the gapping
of the ghost occurs only at very small momenta $\pvecsq \lesssim (\twopiT)^2\,$,
and we are left with the temperature regime \eq{eq:0513}, in which a weakening of the
infrared ghost dominance is to be expected. This structure is compatible with the
results in \cite{Fister:2014bpa}, where no zero crossing was observed 
at temperatures about $T_c$ in the accessible momentum regime.
In our opinion, it would therefore be interesting 
to extend the analysis of \cite{Fister:2014bpa} to smaller momenta. 

\section{Summary \& Outlook}
\label{sec:Conclusion}

We have presented non-perturbative first-principles results for the
finite-temperature Landau-gauge Yang-Mills correlation functions,
obtained from the functional renormalisation group. Our comprehensive
truncation of the effective action includes the computationally
especially expensive magnetic and electric components of the purely
gluonic vertices.
We gauged our truncation by comparing to propagator
results obtained in lattice simulations and found very good agreement
for the magnetic gluon propagator. Our result for the Debye screening
mass shows excellent agreement with two-loop hard thermal loop
perturbation theory at high temperatures and the electric gluon
propagator compares very well to lattice results for all temperatures
except $T\in (0.5 \, T_c\,,\,1.3\, T_c)\,$.
We have argued that the
deviations in this regime are related to the different backgrounds
used.
Particular focus was also put on the fate
of the zero crossing in the three-gluon vertex at finite
temperature. In the electric component of the three-gluon vertex we
found the disappearance of the zero crossing at a very small
temperature.  The magnetic zero crossing also moves towards lower
momenta for small temperatures but it never vanishes.  At high
temperatures, its position increases linearly with the temperature.
We gave an analytic argument for the observed qualitative behaviour of
the zero crossing in the magnetic and electric components.

The presented first-principles results for the finite-temperature
correlation functions of Yang-Mills theory form the foundation for a
number of subsequent studies. First and foremost, the capability to
perform non-perturbative first-principles studies of gauge theories at
finite temperatures provides a crucial prerequisite for the
investigation of the QCD phase structure. In particular, combining the
advancements of this work with those of a recent calculation of the
correlators of two-flavour QCD~\cite{Cyrol:2017ewj}, will allow us to
investigate the properties of quantum chromodynamics at finite
temperature and density from first principles. Furthermore, the
presented correlators can be used to compute thermodynamic quantities
like the pressure, the shear viscosity, as well as the Polyakov loop
potential and spectral functions, the latter being notoriously
difficult to obtain. Additionally, it is suggestive to use the
remarkable agreement of fully non-perturbative results with resummed 
perturbative results, in particular in the Debye mass, to devise 
functionally  assisted analytic applications for the transport and kinetic 
regime in heavy ion collisions. Finally, we expect that improving the current
investigation by including a non-vanishing background field and
non-vanishing Matsubara modes will lead to the disappearance of the
discrepancy in the electric gluon propagator near the phase transition
temperature.

\acknowledgments

We thank Lukas Corell, Ouraman Hajizadeh, Markus Q. Huber, Axel Maas, Richard Williams,
and Nicolas Wink for discussions. This work is supported by EMMI, the
grant ERC-AdG-290623, the FWF through Erwin-Schr\"odinger-Stipendium
No.\ J3507-N27, the Studienstiftung des deutschen Volkes, the DFG
through grant STR 1462/1-1, the BMBF grant 05P12VHCTG, and is part of
and supported by the DFG Collaborative Research Centre "SFB 1225
(ISOQUANT)". It is also supported in part by the Office of Nuclear
Physics in the US Department of Energy's Office of Science under
Contract No. DE-AC02-05CH11231.

\appendix

\section{Regulator and truncation dependence}
\label{app:RegulatorIndependence}

\begin{figure}
	\includegraphics[width=\textwidth]{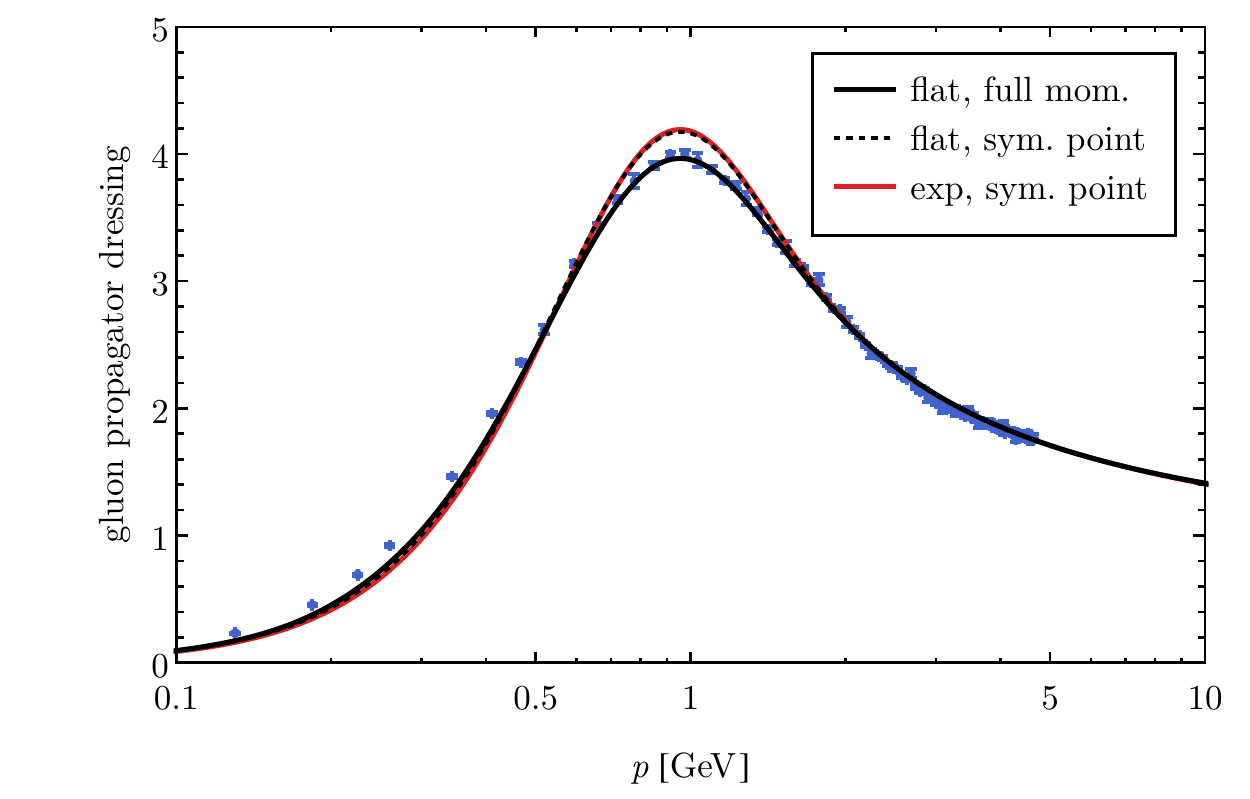}
	\caption{
		Gluon propagator dressing obtained with the exponential regulator 
		in comparison with dressings calculated with the flat regulator in 
		\cite{Cyrol:2016tym} and $SU(3)$ lattice data~\cite{Sternbeck:2006cg}.
		The lattice results are renormalised as in \cite{Cyrol:2016tym}.
		Newer lattice results \cite{Duarte:2016iko} 
		agree with \cite{Sternbeck:2006cg} if the largest 
		physical volumes are compared.\myhfill
	}
	\label{fig:RegulatorIndependence}
\end{figure}

The regulators in \eq{eq:dSk} are parametrised by
\begin{align}
	\label{eq:regulators}
	& R^{ab}_{k,\mu\nu}(p) = \tilde Z^\text{M}_{A,k}\, r(p^2/k^2)\, p^2\, 
		\delta^{ab}\, \pT[p]{\mu\nu}\,,\eqnewline
	& R^{ab}_k(p) = \tilde Z_{c,k}\, r(p^2/k^2)\, p^2 \delta^{ab}\,,
\end{align}
where we dress the regulators with $\tilde Z^\text{M}_{A,k}$ and $\tilde Z_{c,k}$ 
as in~\cite{Cyrol:2016tym}.
Since $\pT[p]{\mu\nu}=\pM[p]{\mu\nu}+\pE[p]{\mu\nu}\,$, \eq{eq:regulators}
implies the same regularisation for electric and magnetic modes. 
Due to its advantages for the evaluation of thermodynamic quantities
\cite{Fister:2011uw,Fister:Diss} we use the exponential 
regulator shape function,
\begin{align}
	\label{eq:shapeFunction}
	r(x)=\frac{x^{m-1}\,e^{-x^m}}{1-e^{-x^m}}\,,
\end{align}
with $m=2\,$. This is in contrast to the vacuum calculations in \cite{Cyrol:2016tym},
which were performed with a smoothed version of the flat regulator \cite{Litim:2000ci}.

In \Fig{fig:RegulatorIndependence}, we show vacuum results obtained with
the flat and the exponential regulator. Clearly, the results obtained with
the symmetric-point approximation, defined by \eq{eq:generalMomentumApprox}, agree very well.
However, they show a higher bump than the lattice results. 
This is due to the symmetric-point approximation used for the vertices.
This discrepancy vanishes if more momentum dependencies are included
as shown in \Fig{fig:RegulatorIndependence}, \cf \cite{Cyrol:2016tym} for a thorough discussion. 
An extension of the current finite-temperature investigations
beyond the symmetric-point approximation is deferred to future work.

\section{Tensor Splitting}
\label{app:TensorSplitting}

At vanishing Matsubara frequencies, not all tensors that are obtained by
contracting the classical tensor structures with all possible 
combinations of electric and magnetic projectors are linearly independent, 
see \App{app:Degeneracy}. Since we calculate the dressing functions at
vanishing Matsubara mode, we can compute only a restricted set of dressings.
This implies that we have to approximate the remaining degenerate dressing 
functions from this reduced set of dressings in order to obtain the correct 
UV behaviour and to recover the vacuum results in the zero-temperature limit.
We use
\begin{align}
	\label{eq:GhostGluonVertex}
	& [\Gamma^{(3)}_{A\bar{c}c}]^{abc}_{\mu} (p,q) = 
		[S^{(3)}_{A\bar{c}c}]^{abc}_{\mu'} (p,q) \,  \eqnewline
	& \quad \times \Big( \lAccM(\psym) \; \pM{\mu'\mu} + \lAccM(\psym) \; 
		\pE{\mu'\mu}\Big)\,
\end{align}
for the ghost-gluon vertex,
\begin{align}
	\label{eq:ThreeGluonVertex}
	[\Gamma^{(3)}_{A^3}]^{abc}_{\mu\nu\rho} & (p,q) = 
		[S^{(3)}_{A^3}]^{abc}_{\mu'\nu'\rho'} (p,q) \cdot  \eqnewline
			\Big(
				& \lAAAM(\psym) \; \pM{\mu'\mu} \, \pM{\nu'\nu} \, \pM{\rho'\rho} \, +
				\eqnewline
				& \lAAAM(\psym) \; \pE{\mu'\mu} \, \pM{\nu'\nu} \, \pM{\rho'\rho} \, + 
				\perm \, +\eqnewline
				& \lAAAE(\psym) \; \pE{\mu'\mu} \, \pE{\nu'\nu} \, \pM{\rho'\rho} \, + 
				\perm \, +\eqnewline
				& \lAAAM(\psym) \; \pE{\mu'\mu} \, \pE{\nu'\nu} \, \pE{\rho'\rho} \; 
			\Big)\,,
\end{align}
for the three-gluon vertex and
\begin{align}
	\label{eq:FourGluonVertex}
	[\Gamma^{(4)}_{A^4}]^{abcd}_{\mu\nu\rho\sigma} & (p,q,r) = 
		[S^{(4)}_{A^4}]^{abcd}_{\mu'\nu'\rho'\sigma'}\, \cdot  \eqnewline
		\Big(
			& \lAAAAM(\psym) \; \pM{\mu'\mu} \, \pM{\nu'\nu} \, \pM{\rho'\rho} \, \pM{\sigma'\sigma} \, +\eqnewline
			& \lAAAAM(\psym) \; \pE{\mu'\mu} \, \pM{\nu'\nu} \, \pM{\rho'\rho} 
			\, \pM{\sigma'\sigma} \, + \perm \, +\eqnewline
			& \lAAAAE(\psym) \; \pE{\mu'\mu} \, \pE{\nu'\nu} \, \pM{\rho'\rho} 
			\, \pM{\sigma'\sigma} \, + \perm \, +\eqnewline
			& \lAAAAM(\psym) \; \pE{\mu'\mu} \, \pE{\nu'\nu} \, \pE{\rho'\rho} 
			\, \pM{\sigma'\sigma} \, + \perm \, +\eqnewline
			& \lAAAAM(\psym) \; \pE{\mu'\mu} \, \pE{\nu'\nu} \, \pE{\rho'\rho} 
			\, \pE{\sigma'\sigma} \; 
		\Big)\,,
\end{align}
for the four-gluon vertex.
Here and in the following we leave the momentum arguments implicit.
Although the dressing functions of some tensors coincide in our
approximation, we explicitly show the splitting to make the construction of the approximation apparent. 
The magnetic dressing functions
appear in more than one tensor structure, and we evaluate them by projecting
onto the purely magnetic tensors structure for every vertex, see \App{app:Projections}.
Due to the $O(4)$-symmetry of the vacuum,
this approximation becomes exact for large momenta $p^2 \gg (\twopiT)^2\,$, which are 
not affected by finite-temperature effects. 
In the limit of vanishing Matsubara frequencies, the dimension
of the tensor space is reduced, see \App{app:Degeneracy}.
Therefore, this
approximation is very good also for small momenta $p^2 \lesssim (\twopiT)^2\,$.
Hence, the approximations used in \eqrange{eq:GhostGluonVertex}{eq:FourGluonVertex} 
affect only intermediate Matsubara modes,
which are only slightly influenced by finite temperature effects,
see \Sec{sec:Setup:VertexExpansion}.

\begin{figure}
	\includegraphics[width=\textwidth]{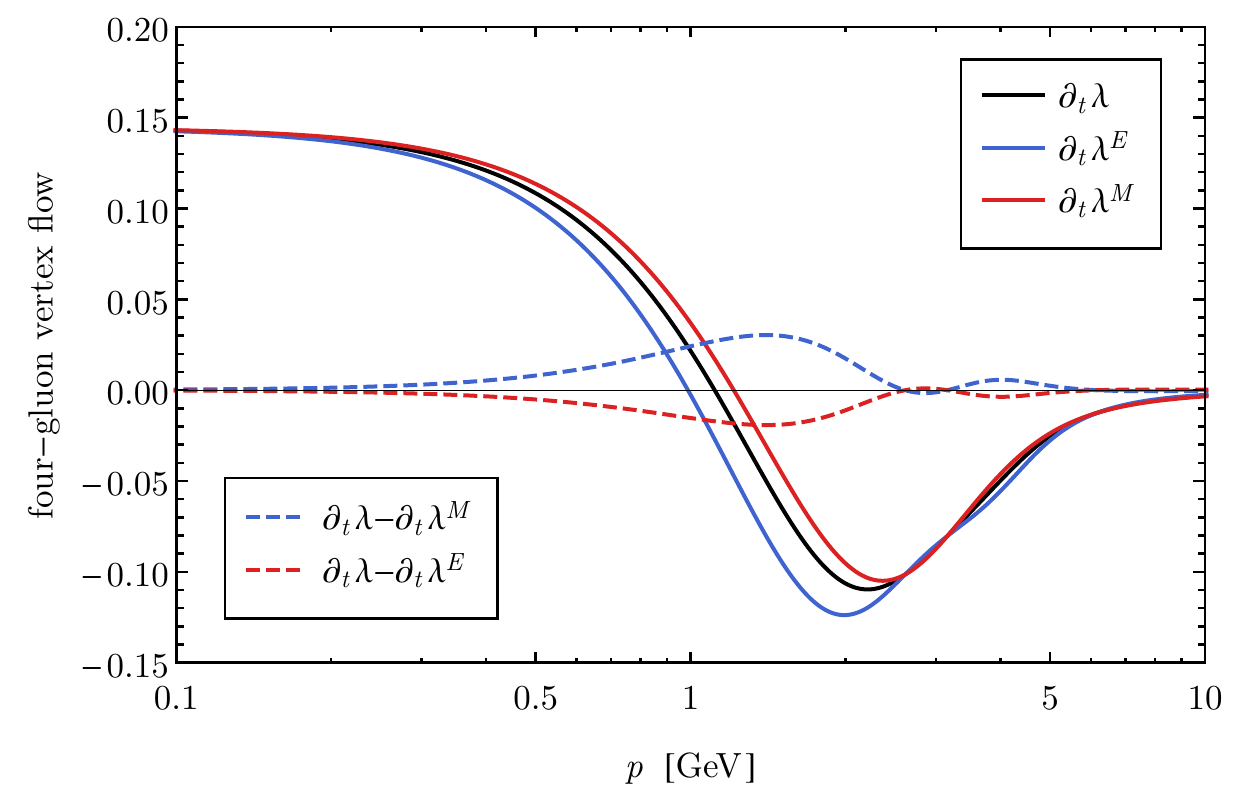}
	\caption{
		Four-gluon vertex vacuum flows from different projection operators, 
		\eq{eq:FourGluonVertexVacuumProjection} and \eq{eq:FourGluonVertexProjection}, and their differences
		 at the RG scale $k=\SI{2}{\GeV}$. \myhfill
	}
	\label{fig:FourGluonVertexProjection}
\end{figure}

\begin{figure*}
	\ffigbox{
		\begin{subfloatrow}
			\ffigbox{\includegraphics{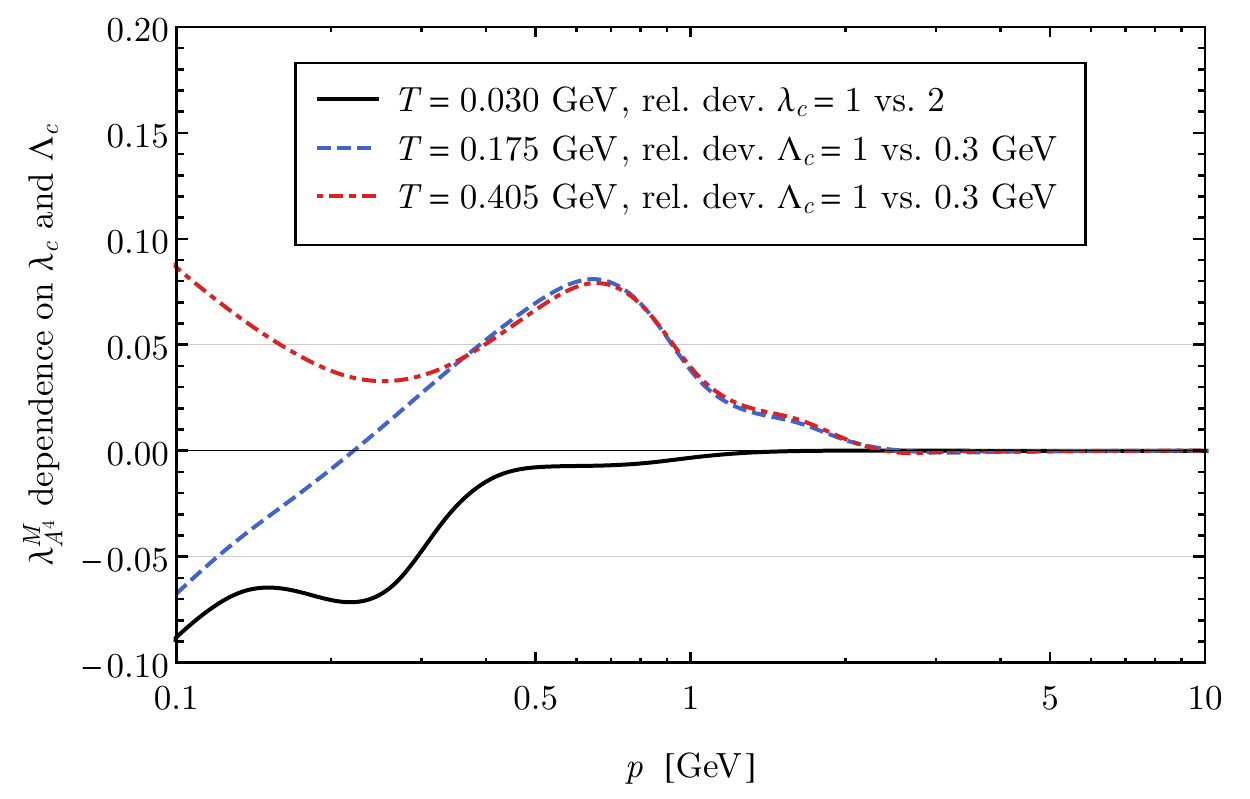}}{
				\caption{Magnetic dressing.}
				\label{fig:CorrectionScales:Magnetic}
			} 
			\ffigbox{\includegraphics{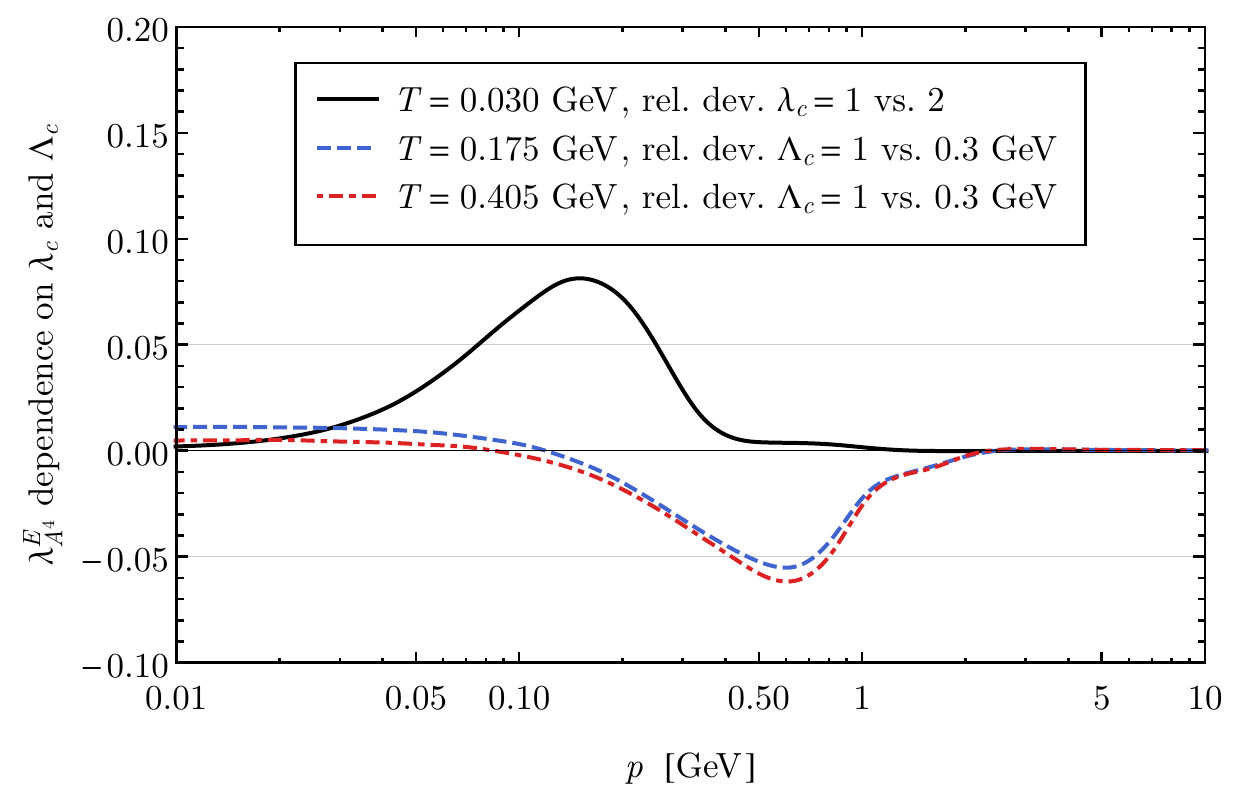}}{
				\caption{Electric dressing.}
				\label{fig:CorrectionScales:Electric}
			}
		\end{subfloatrow}
	}{
	\caption{
		Relative deviations, \eg $\left(\lAAAAM(\lambda_c=1)-\lAAAAM(\lambda_c=2)\right)/\lAAAAM(\lambda_c=1)\,$, of the four-gluon 
		vertex dressings, \eq{eq:FourGluonVertex}, calculated with different parameters in the smoothed theta function \eq{eq:ThetaFunction}.
		Depending on the temperature, the dressings depend either on $\lambda_c$ or $\Lambda_c\,$, see \eq{eq:ThetaScale}.\myhfill
	}
	\label{fig:CorrectionScales}
}
\end{figure*}

\section{Tensor Degeneracy}
\label{app:Degeneracy}

We generalise the classical tensor structures to finite temperature by attaching 
all combinations of magnetic and electric projection operators, 
see \eqrange{eq:GhostGluonVertex}{eq:FourGluonVertex}.
However, contracting the electric ghost-gluon vertex with itself and evaluating it at 
vanishing Matsubara modes yields
\begin{align}
	\label{eq:GhostGluonVertexReduction}
	[S^{(3)}_{A\bar{c}c}]^{abc}_{\mu} \; \pE{\mu\mu'} \; 
		[S^{(3)}_{A\bar{c}c}]^{abc}_{\mu'} \Big|_{\{n_i=0\}} = 0\,.
\end{align}
Hence, the electric component of the ghost-gluon vertex disappears 
in the limit of vanishing Matsubara frequencies.
Similarly, we find for the three-gluon vertex,
\begin{align}
	\label{eq:ThreeGluonVertexReduction}
	& [S^{(3)}_{A^3}]^{abc}_{\mu\nu\rho} \;
		\pE{\mu\mu'} \, \pM{\nu\nu'} \, \pM{\rho\rho'} \;
		[S^{(3)}_{A^3}]^{abc}_{\mu'\nu'\rho'}
		\Big|_{\{n_i=0\}} = 0\, , \eqnewline
	& [S^{(3)}_{A^3}]^{abc}_{\mu\nu\rho} \; 
		\pE{\mu\mu'} \, \pE{\nu\nu'} \, \pE{\rho\rho'} \;
		[S^{(3)}_{A^3}]^{abc}_{\mu'\nu'\rho'}
		\Big|_{\{n_i=0\}} = 0\,,
\end{align}
and for the four-gluon vertex,
\begin{align}
	\label{eq:FourGluonVertexReduction}
	[S^{(4)}_{A^4}]^{abcd}_{\mu\nu\rho\sigma} \; \pE{\mu\mu'} \, 
		\pM{\nu\nu'} \, \pM{\rho\rho'} \, \pM{\sigma\sigma'} 
		[S^{(4)}_{A^4}]^{abcd}_{\mu'\nu'\rho'\sigma'} 
		\Big|_{\{n_i=0\}} &= 0\,,\eqnewline
	[S^{(4)}_{A^4}]^{abcd}_{\mu\nu\rho\sigma} \; \pE{\mu\mu'} \, 
		\pE{\nu\nu'} \, \pE{\rho\rho'} \, \pM{\sigma\sigma'} 
		[S^{(4)}_{A^4}]^{abcd}_{\mu'\nu'\rho'\sigma'}
		\Big|_{\{n_i=0\}} &= 0\,,\eqnewline
	[S^{(4)}_{A^4}]^{abcd}_{\mu\nu\rho\sigma} \; \pE{\mu\mu'} \, 
		\pE{\nu\nu'} \, \pE{\rho\rho'} \, \pE{\sigma\sigma'} 
		[S^{(4)}_{A^4}]^{abcd}_{\mu'\nu'\rho'\sigma'}
		\Big|_{\{n_i=0\}} &= 0\,.
\end{align}
Thus, for $p^2 \ll (\twopiT)^2$ the classical vertex dressings are fully 
described by the remaining basis tensors, to wit, 
those with only magnetic legs and those with exactly two electric legs.

\section{Projecting the flow equations}
\label{app:Projections}

The tensor bases for the propagators as well as for the ghost-gluon
vertex are complete, and therefore the projection onto the dressings is
unique. For the gluonic vertices we do not take the full transverse
tensor bases into account.  Consequently, already in the vacuum, any
projection is an approximation that relies on the assumption that
non-included basis elements are small.  If the flows are projected
onto their electric and magnetic components, the incompleteness of the
bases can lead to intricate complications. The reason is that the
magnetic and electric projection operators can yield differing
contributions from non-classical tensor structures that are created by
quantum fluctuations. As an immediate consequence, the magnetic and
electric dressings differ then by momentum dependent terms. This
effect occurs already at vanishing temperature, and is therefore in
contradiction with the $O(4)$-symmetry of the vacuum.  If one uses a
complete basis, projecting with magnetic and
electric projection operators does not spoil the $O(4)$-symmetry
although the projection operators themselves are not $O(4)$-symmetric.

In the following two subsections we discuss in detail the quantitative
relevance of these effects caused by the incomplete bases for the
gluonic vertices.  In order to disentangle genuine finite-temperature
contributions from these projection artefacts, we consider only vacuum
flows. By splitting the projection into electric and magnetic
components and comparing them to the $O(4)$-symmetric projection, we
are able to quantify these basis artefacts. Unfortunately, we find
that the emergence of certain non-classical tensors yields sizeable
artefacts on the dressing of the classical tensor structure of the
four-gluon vertex. As discussed in detail in this and the following
two \Apps{app:VacuumLimit}{app:StartScale},
implementing a proper treatment of these artefacts of the incomplete
bases turns out to be vital to obtain the correct UV behaviour and
cutoff independence of the finite-temperature results.

\subsection{Three-gluon vertex}
\label{app:Projections:ThreeGluonVertex}
We project onto the magnetic and electric components of the three-gluon vertex by
\begin{align}
	\label{eq:ThreeGluonVertexProjection}
	\lAAAM &= 
		\frac{[S^{(3)}_{A^3}]^{abc}_{\mu\nu\rho} \; \pM{\mu\mu'} \, \pM{\nu\nu'} \, \pM{\rho\rho'} \;  [\Gamma^{(3)}_{A^3}]^{abc}_{\mu'\nu'\rho'}}{[S^{(3)}_{A^3}]^{abc}_{\mu\nu\rho} \; \pM{\mu\mu'} \, \pM{\nu\nu'} \, \pM{\rho\rho'} \; [S^{(3)}_{A^3}]^{abc}_{\mu'\nu'\rho'}}\,,\eqnewline
	\lAAAE &= 
		\frac{[S^{(3)}_{A^3}]^{abc}_{\mu\nu\rho} \; \pE{\mu\mu'} \, \pE{\nu\nu'} \, \pM{\rho\rho'} \;  [\Gamma^{(3)}_{A^3}]^{abc}_{\mu'\nu'\rho'}}{[S^{(3)}_{A^3}]^{abc}_{\mu\nu\rho} \; \pE{\mu\mu'} \, \pE{\nu\nu'} \, \pM{\rho\rho'} \; [S^{(3)}_{A^3}]^{abc}_{\mu'\nu'\rho'}}\,,
\end{align}
as generalisation of the vacuum projection
\begin{align}
	\label{eq:ThreeGluonVertexVacuumProjection}
	\lAAA &= 
		\frac{[S^{(3)}_{A^3}]^{abc}_{\mu\nu\rho} \; \pT{\mu\mu'} \, \pT{\nu\nu'} \, \pT{\rho\rho'} \;  [\Gamma^{(3)}_{A^3}]^{abc}_{\mu'\nu'\rho'}}{[S^{(3)}_{A^3}]^{abc}_{\mu\nu\rho} \; \pT{\mu\mu'} \, \pT{\nu\nu'} \, \pT{\rho\rho'} \; [S^{(3)}_{A^3}]^{abc}_{\mu'\nu'\rho'}}\,.
\end{align}
In explicit numerical checks we find that the projections \eq{eq:ThreeGluonVertexProjection} and 
\eq{eq:ThreeGluonVertexVacuumProjection} agree to the per mille level at $T=0$ and therefore also for 
$k \gg \twopiT\,$. We conclude that our projection is not sensitive to the possible emergence of non-classical
tensors structures in the three-gluon vertex. This is also consistent with the sub-leading importance of non-classical 
tensor structures found in earlier three-gluon vertex studies \cite{Eichmann:2014xya}.

\begin{figure*}
	\ffigbox{
		\begin{subfloatrow}
			\ffigbox{\includegraphics{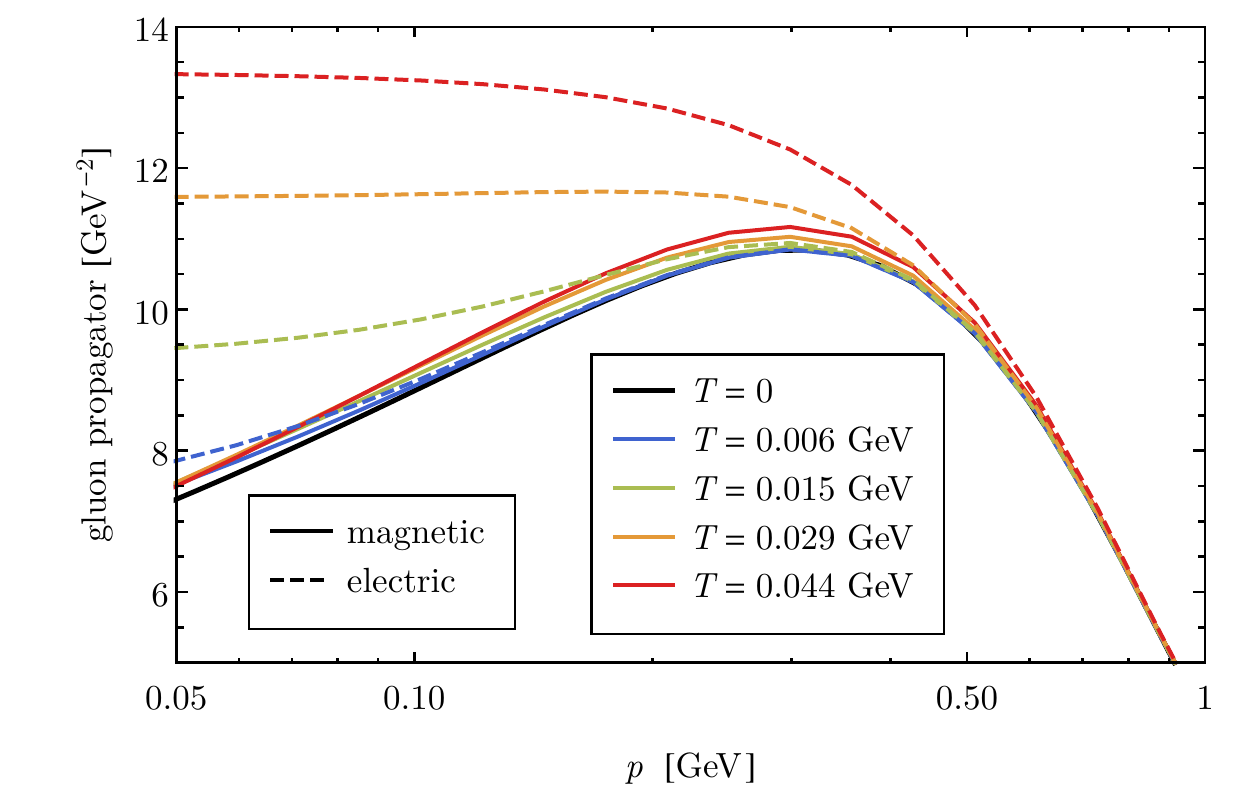}}{
				\caption{Convergence towards the vacuum results. The plot shows the degeneracy of the magnetic and the electric gluon propagators for $p \gtrsim \twopiT\,$.\myhfill}
				\label{fig:VacuumLimit}
			} 
			\ffigbox{\includegraphics{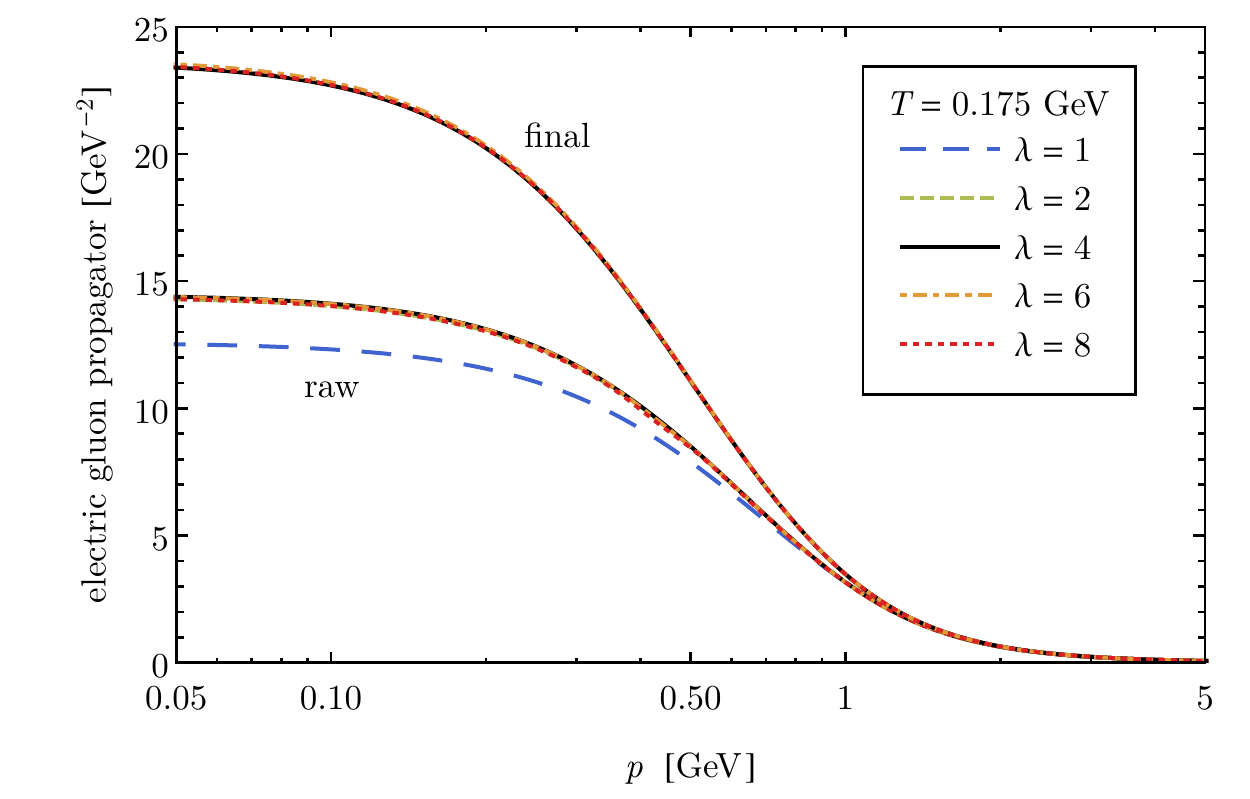}}{
				\caption{Gluon propagator obtained with different initial scales $\Lambda_T = \lambda \, \twopiT\,$.
					See \App{app:UVmass} for the definition of raw and final.
					We do not show the final propagator for $\lambda=1$ since in this case the start scale $\Lambda_T=\twopiT$ lies below the correction scale $k_T=\fourpiT\,$, \cf \eq{eq:GluonPropagotorCorrection}.\myhfill}
				\label{fig:StartScale}
			}
		\end{subfloatrow}
	}{
	\caption{
		Vacuum limit and initial scale independence of gluon propagator, \eq{eq:propagators}.
	}
	\label{fig:vacuumlimitandstartscale}
}
\end{figure*}

\subsection{Four-gluon vertex}
\label{app:Projections:FourGluonVertex}

We project onto the vacuum dressing function with
\begin{align}
	\label{eq:FourGluonVertexVacuumProjection}
	\lAAAA = 
	\frac{[S^{(4)}_{A^4}]^{abcd}_{\mu\nu\rho\sigma} \; \pT{\mu\mu'} \, \pT{\nu\nu'} \, \pT{\rho\rho'} \, \pT{\sigma\sigma'} \;  [\Gamma^{(4)}_{A^4}]^{abcd}_{\mu'\nu'\rho'\sigma'}}{[S^{(4)}_{A^4}]^{abcd}_{\mu\nu\rho\sigma} \; \pT{\mu\mu'} \, \pT{\nu\nu'} \, \pT{\rho\rho'} \, \pT{\sigma\sigma'} \; [S^{(4)}_{A^4}]^{abcd}_{\mu'\nu'\rho'\sigma'}}\,.
\end{align}
Assuming vanishing non-classical tensor structures, this generalises to
\begin{align}
	\label{eq:FourGluonVertexProjection}
	\lAAAAM &= 
	\frac{[S^{(4)}_{A^4}]^{abcd}_{\mu\nu\rho\sigma} \; \pM{\mu\mu'} \, \pM{\nu\nu'} \, \pM{\rho\rho'} \, \pM{\sigma\sigma'} \;  [\Gamma^{(4)}_{A^4}]^{abcd}_{\mu'\nu'\rho'\sigma'}}{[S^{(4)}_{A^4}]^{abcd}_{\mu\nu\rho\sigma} \; \pM{\mu\mu'} \, \pM{\nu\nu'} \, \pM{\rho\rho'} \, \pM{\sigma\sigma'} \; [S^{(4)}_{A^4}]^{abcd}_{\mu'\nu'\rho'\sigma'}}\,,\eqnewline
	\lAAAAE &= 
	\frac{[S^{(4)}_{A^4}]^{abcd}_{\mu\nu\rho\sigma} \; \pE{\mu\mu'} \, \pE{\nu\nu'} \, \pM{\rho\rho'} \, \pM{\sigma\sigma'} \;  [\Gamma^{(4)}_{A^4}]^{abcd}_{\mu'\nu'\rho'\sigma'}}{[S^{(4)}_{A^4}]^{abcd}_{\mu\nu\rho\sigma} \; \pE{\mu\mu'} \, \pE{\nu\nu'} \, \pM{\rho\rho'} \, \pM{\sigma\sigma'} \; [S^{(4)}_{A^4}]^{abcd}_{\mu'\nu'\rho'\sigma'}}\,,
\end{align}
for the magnetic and the electric components.
If the only tensor generated by the flow were the classical one,
\begin{align}
	[\Gamma^{(4)}_{A^4}]^{abcd}_{\mu\nu\rho\sigma}\propto
		[S^{(4)}_{A^4}]^{abcd}_{\mu\nu\rho\sigma}=
		f^{abn}f^{cdn}\delta_{\mu\rho}\delta_{\nu\sigma} + \perm\,,
\end{align}
the projections \eq{eq:FourGluonVertexVacuumProjection} and \eq{eq:FourGluonVertexProjection} would 
yield $\lAAAA=\lAAAAM=\lAAAAE$. However, this equality can be spoiled by the presence of non-classical
tensors, which are in general created by the flow equation. Consider, for example, the 
following $O(4)$- and Bose-symmetric non-classical tensor:
\begin{align}
	\label{eq:BadNonClassicalTensor}
	&[\Gamma^{(4)}_{A^4,\text{ncl}}]^{abcd}_{\mu\nu\rho\sigma}(p,q,r,s) = \big(
		f^{abn}f^{cdn}\, \cdot \eqnewline
		&\qquad (q+s)_\mu (q+s)_\rho
				(p+r)_\nu (p+r)_\sigma
				\big) + \perm\,.
\end{align}
Inserting \eq{eq:BadNonClassicalTensor} into \eq{eq:FourGluonVertexVacuumProjection} and 
\eq{eq:FourGluonVertexProjection} yields differing contributions to the dressing functions $\lAAAA\,$, $\lAAAAM\,$, and $\lAAAAE\,$.
Therefore, $O(4)$-invariance is lost
due to the incompleteness of the basis that was used to construct the projection operators, \eq{eq:FourGluonVertexVacuumProjection} and \eq{eq:FourGluonVertexProjection}.

In \Fig{fig:FourGluonVertexProjection} we show the vacuum flows of the four-gluon vertex obtained with different 
projection operators and identical vacuum vertices on the right hand side of the flow equation.
In contrast to the three-gluon vertex, we find a considerable difference in the resulting momentum dependence
of the projections \eq{eq:FourGluonVertexVacuumProjection} and \eq{eq:FourGluonVertexProjection}.
We conclude, that sizeable non-classical tensors, which affect the difference between
the magnetic and electric projection operators, are generated. As an immediate consequence, the $O(4)$-symmetric limit at $T\rightarrow 0$ 
is spoiled by the presence of these tensors since $\lAAAAM(T=0)\neq\lAAAAE(T=0)\,$.

\begin{figure*}
	\ffigbox{
		\begin{subfloatrow}
			\ffigbox{\includegraphics{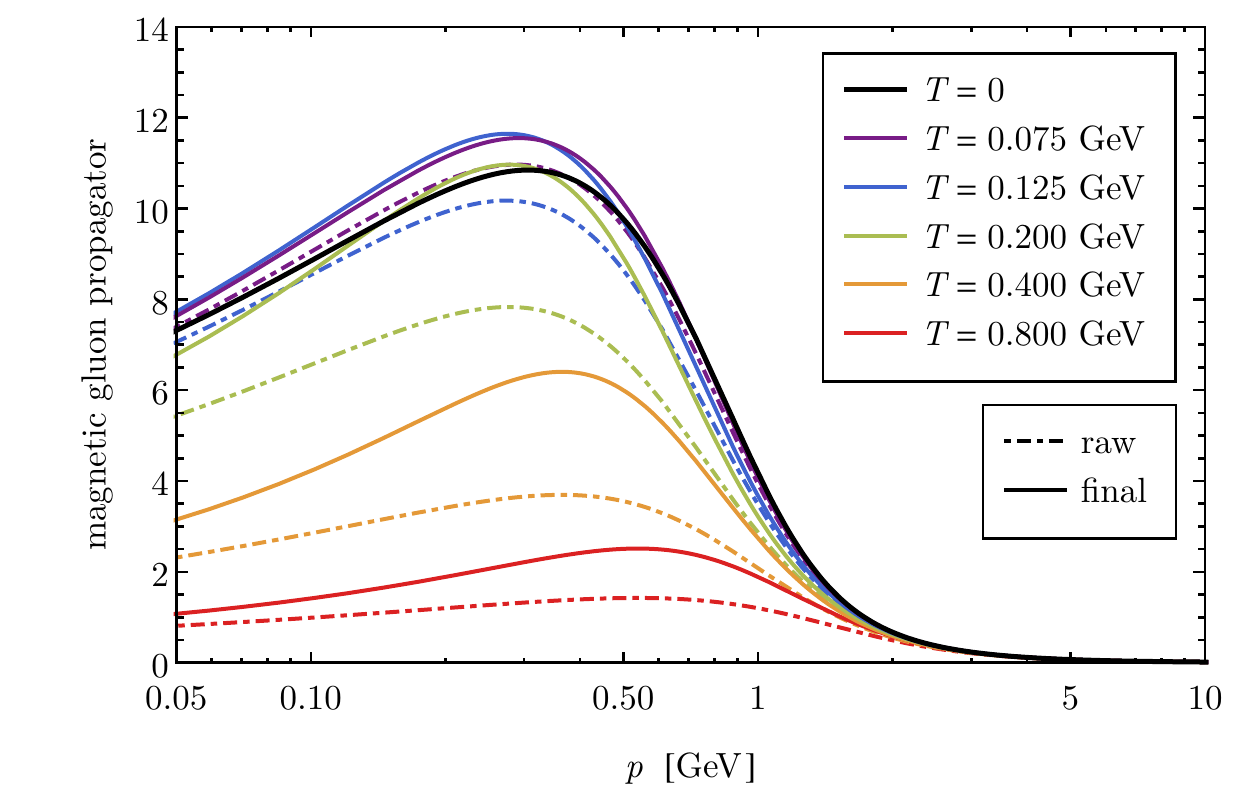}}{
				\caption{Magnetic gluon propagator $1/\left(p^2\ZAM(p)\right)\,$.}
			} 
			\ffigbox{\includegraphics{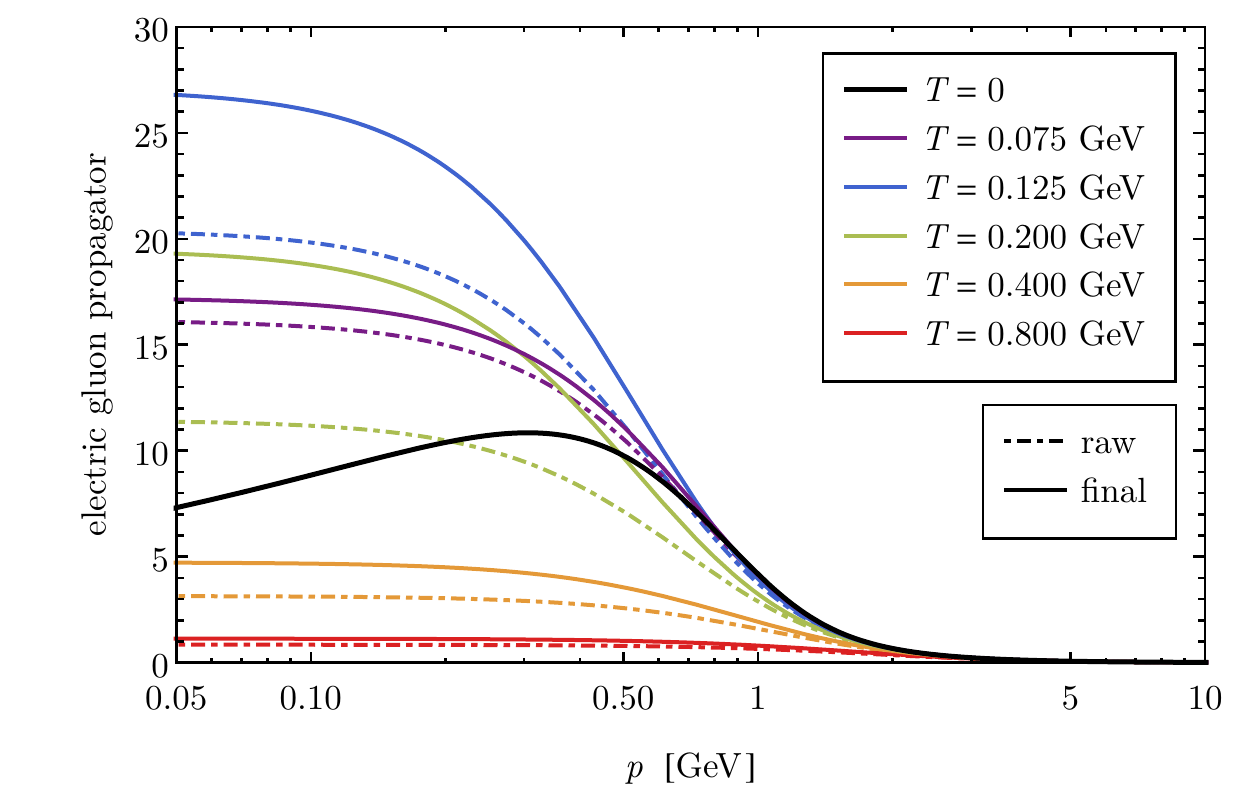}}{
				\caption{Electric gluon propagator $1/\left(p^2\ZAE(p)\right)\,$.}
			}
		\end{subfloatrow}
	}{
	\caption{Electric and magnetic gluon propagators with and without mass subtraction procedure, \eq{eq:GluonPropagotorCorrection}.}
	\label{fig:EffectOfSubtraction}
}
\end{figure*}

A simple estimate of the unphysical projection artefacts of these non-classical tensors is given by the vacuum differences
of the projections \eqs{eq:FourGluonVertexVacuumProjection} and \eq{eq:FourGluonVertexProjection},
\begin{align}
	\label{eq:fourGluonVertexProjectionCorrection_0}
	\partial_t \left[\lAAAA - \lAAAAM(T=0)\right]\,,\eqnewline
	\partial_t \left[\lAAAA - \lAAAAE(T=0)\right]\,,
\end{align}
which are also shown in \Fig{fig:FourGluonVertexProjection}.
Assuming that this unphysical difference depends only mildly on the temperature,
a natural strategy to account for this artefact is to subtract \eq{eq:fourGluonVertexProjectionCorrection_0} from the
finite-temperature flows. However, there is an additional complication in the case of the scaling solution.
The vertex dressings obey a power law behaviour at small momenta, see  \eq{eq:GeneralScaling}, and the corresponding exponent changes as one goes from the vacuum to finite temperature.
Although this behaviour of the correlators at very small momenta
does not affect any observables, it has to be taken into account when subtracting \eq{eq:fourGluonVertexProjectionCorrection_0}
from the finite-temperature flows.
Consequently, we modify the flows of the magnetic and electric components by 
\begin{align}
	\label{eq:fourGluonVertexProjectionCorrection}
	\partial_t \lAAAAM(T) = \partial_t \lAAAAM(T) + \theta_\epsilon(k,\,k_c) \, \partial_t\left[\lAAAA - \lAAAAM(T=0)\right],\eqnewline
	\partial_t \lAAAAE(T) = \partial_t \lAAAAE(T) + \theta_\epsilon(k,\,k_c) \, \partial_t\left[\lAAAA - \lAAAAE(T=0)\right].
\end{align}
The purpose of the smoothed step function,
\begin{align}
	\label{eq:ThetaFunction}
	\theta_\epsilon(k,\,k_c)=\frac{1}{1+\exp\left[\frac{1}{\epsilon}\left(1-\frac{k}{k_c}\right)\right]}\,,
\end{align}
is to provide a transition from the corrected flows to
the pure finite-temperature flows 
with the correct scaling behaviour at very low momenta. We set the transition scale $k_c$ to
\begin{align}
	\label{eq:ThetaScale}
	k_c = \min\left(\lambda_c\,\twopiT,\,\Lambda_c\right)\,,
\end{align}
which is defined in terms of the parameters $\Lambda_c$ and $\lambda_c\,$.
The modified dressings fulfill
\begin{align}
	\label{eq:VacuumLimit}
	\lim_{T \rightarrow 0} \lAAAAM(T) = \lim_{T \rightarrow 0} \lAAAAE(T) =\lAAAA\,.
\end{align}
This guarantees that we recover the vacuum results in the limit of vanishing temperature
while our best estimates for the basis artefacts are subtracted above the transition scale $k_c$.

In order to investigate the influence of the precise value of the
transition scale, we vary $\Lambda_c$ and $\lambda_c$ in reasonable
ranges. Since temperature effects are expected to be small at momentum
scales $k\geq \twopiT\,$, $\lambda_c$ should be of order unity and we
vary it from $1$ to $2\,$. Furthermore, the gapping scale of the gluon
propagator gives us an estimate on the scale below which no
phenomenologically important effects are to be expected. Consequently,
we vary $\Lambda_c$ between the location of the maxima of the gluon
propagator and the gluon propagator dressing, \ie
$\Lambda_c\in[0.3,1]$ \si{\GeV}. We find only a mild (\SI{10}{\percent}) dependence
of the four-gluon vertex dressings on these parameters as shown in
\Fig{fig:CorrectionScales}. Since the four-gluon vertex is the least
important of all classical tensors in the self-consistently coupled
system, we find that the dependence of all other dressings on these
parameters is even smaller. For example, the induced uncertainty on
the electric gluon propagator is at most \SI{3}{\percent}, but for a
wide range of temperatures and momenta it is even smaller than
\SI{0.5}{\percent}.  In all cases the dependence on the smoothing
parameter, $\epsilon=0.05\,$, is negligible.

\section{Vacuum limit}
\label{app:VacuumLimit}

We constructed the truncation such that it converges to the
symmetric-point approximation used in \cite{Cyrol:2016tym} in the
vacuum limit, $T\rightarrow 0\,$.  In \Fig{fig:VacuumLimit}, we show
the gluon propagator for a range of small temperatures.  We clearly
see that the magnetic as well as the electric propagators approach the
vacuum propagator in the zero temperature limit. In particular, for
each temperature there exists a threshold momentum above which the
magnetic, the electric and the vacuum dressings agree.  This is not
only a strong check of our code but also shows the validity of
\eqrange{eq:GhostGluonVertex}{eq:FourGluonVertex} at $k\ll T$ as well
as the consistency of our $O(4)$-symmetric momentum approximations,
\ie \eqs{eq:generalMomentumApprox}.  Similarly, the magnetic and
electric dressing functions of the other vertices become degenerate in
the vacuum limit. As discussed in the next appendix, this behaviour
allows us to significantly reduce the computational effort.

\begin{figure*}
	\ffigbox{
		\begin{subfloatrow}
			\ffigbox{\includegraphics{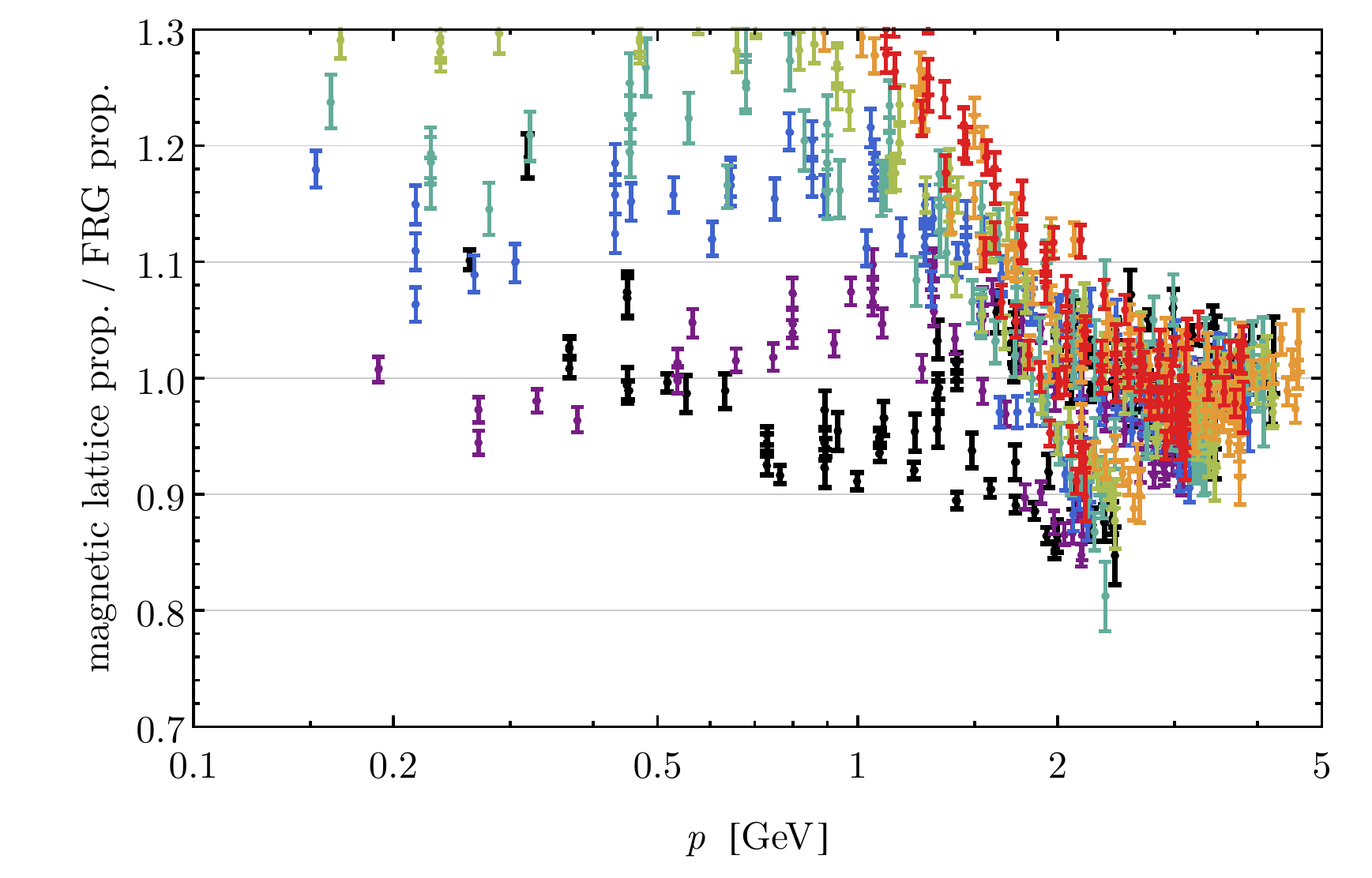}}{
				\caption{Raw propagator.}
				\label{fig:SubtractionLatticeComparison_raw}
			} 
			\ffigbox{\includegraphics{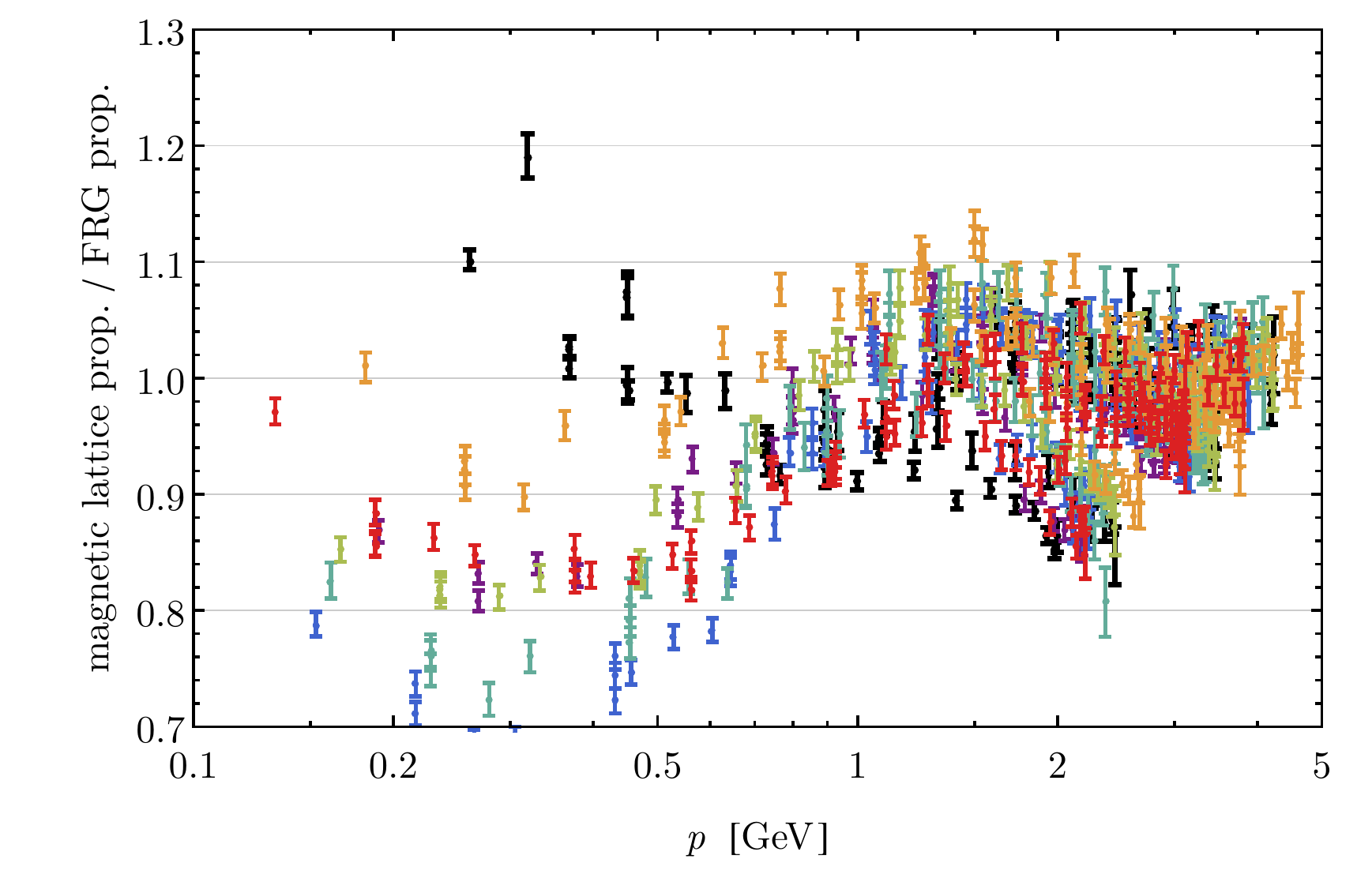}}{
				\caption{Final propagator.}
				\label{fig:SubtractionLatticeComparison_final}
			}
		\end{subfloatrow}
	}{
	\caption{
		Magnetic $SU(2)$ lattice propagator
		\cite{Maas:2011ez,Maas:PC} over FRG propagator, see
		\Fig{fig:MagneticGluonProp:SU2} for the colour coding.
	}
	\label{fig:SubtractionLatticeComparison}
	}
\end{figure*}

\section{Initial scale}
\label{app:StartScale}

The regulator suppresses quantum as well as thermal fluctuations below the regulator scale $k\,$.
Therefore, the temperature-dependent generalised effective action $\Gamma_k(T)$ agrees with the zero-temperature
effective action as long as temperature fluctuations are suppressed,
\begin{align}
	\label{eq:TemperatureSuppression}
	\Gamma_{k}(T)=\Gamma_{k}(T=0) \quad \text{if} \quad \twopiT \ll k\,.
\end{align}
This property enables us to reduce the computational effort by one to two orders of magnitude.
First we compute the $T=0$ effective action, starting at a large 
perturbative scale of typically $k=\Lambda=\SI{60}{\GeV}$ from the classical action.
To obtain the temperature-dependent $1$PI effective action, we integrate
the flow equation starting from the effective average action
$\Gamma_{\Lambda_T}(T)=\Gamma_{\Lambda_T}(T=0)$ at a lower, temperature-dependent cutoff scale,
\begin{align}
	\label{eq:LambdaScale}
	\Lambda_T = \max\left(\lambda\,\twopiT,\,\LambdaTmin\right)\,.
\end{align}
Here, $\LambdaTmin$ has been introduced to avoid the interference of the lowered
starting scale with the dynamical mass generation of the gluon. This is necessary,
because the scaling condition forces us to modify the input at $\Lambda_T$ by a 
gluon mass term, see \App{app:UVmass}. Consequently,
we choose $\LambdaTmin$ as the scale where the vacuum gluon propagator dressing 
becomes maximal, \ie $\LambdaTmin=\SI{0.955}{\GeV}$. We show the dependence 
of the longitudinal gluon propagator on the physical start scale $\lambda$ in 
\Fig{fig:StartScale}. The gluon propagator as well as all
other quantities do not depend on the start scale for $\lambda \geq 2\,$. 
In our numerical computation we use $\lambda=4$ although $\lambda=2$ is sufficient, as argued in \App{app:UVmass}.

To demonstrate the advantage of the temperature-dependent initial scale,
we consider the numerical vacuum integration,
\begin{align}
	\label{eq:VacuumIntegration}
	\int_q = \int \frac{\dd^4 q}{(2\pi)^4} = \int_{0}^{L} 
		\frac{\dd q}{(2\pi)^4} \, q^3 \int \dd\Omega\,.
\end{align}
Numerically, it is advantageous to choose a $k$-dependent numerical cutoff 
$L=l \, k\,$,
where $l=3$ is sufficient for the exponential regulator due to the 
regulator derivative appearing in all diagrams.
This persists in the Matsubara formalism and we can limit the summation to 
frequencies $\omega=\twopiT \, n$ smaller than $L=l k\,$,
\begin{align}
	\label{eq:TemperatureIntegration}
	\sumint_q = \int_{0}^{L} \frac{\dd q}{(2\pi)^3} \, q^2 
		\int \dd\Omega \; T \sum_{n}^{w\leq L}\,.
\end{align}
Thus, the number of required integrand evaluations grows linearly with $k$
as well as with $1/T\,$. For small temperatures, the increased number
of evaluations due to the growing number of small Matsubara modes 
is therefore compensated by the shrinking initial scale, at least
down to $T = \LambdaTmin/(2\pi\lambda)\,$.

\begin{figure*}
	\ffigbox{
		\begin{subfloatrow}
			\ffigbox{\includegraphics{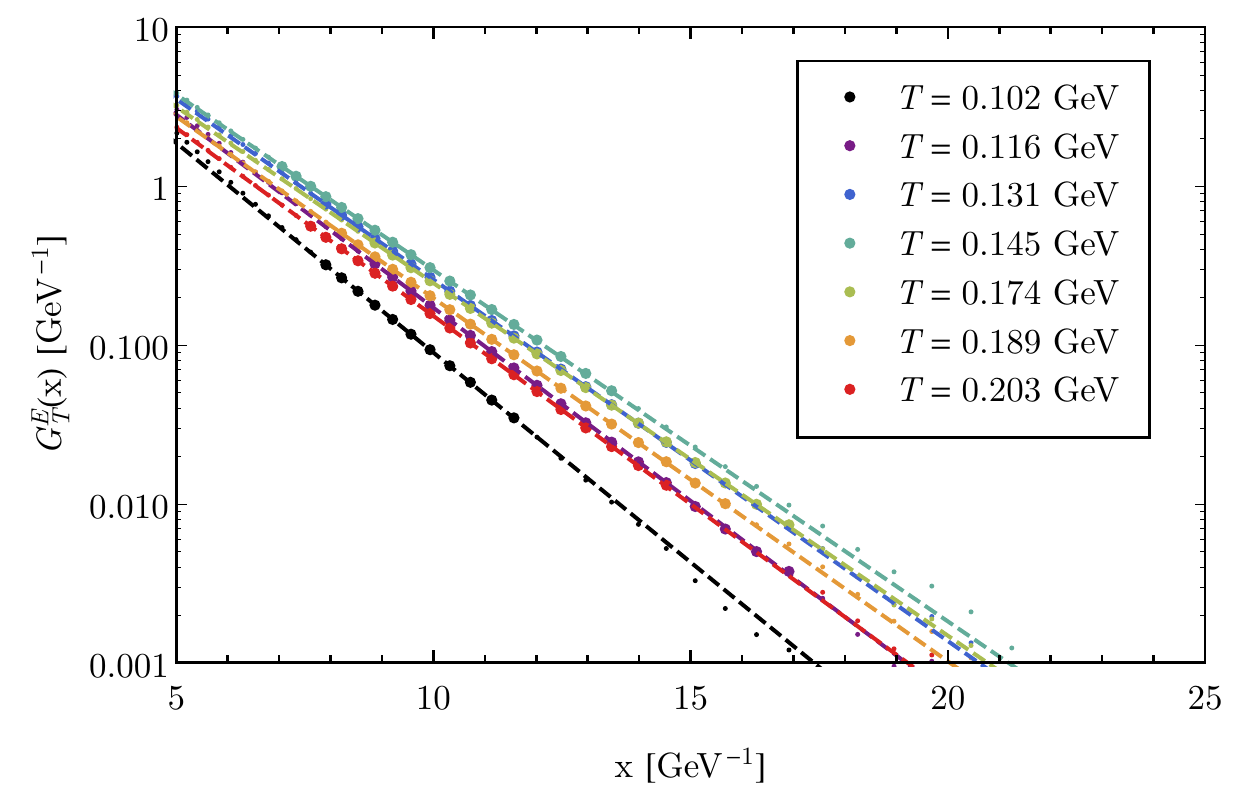}}{
				\caption{$\GE(x)$ for low temperatures.}
				\label{fig:DebyeMassFit:LowT}
			}
			\ffigbox{\includegraphics{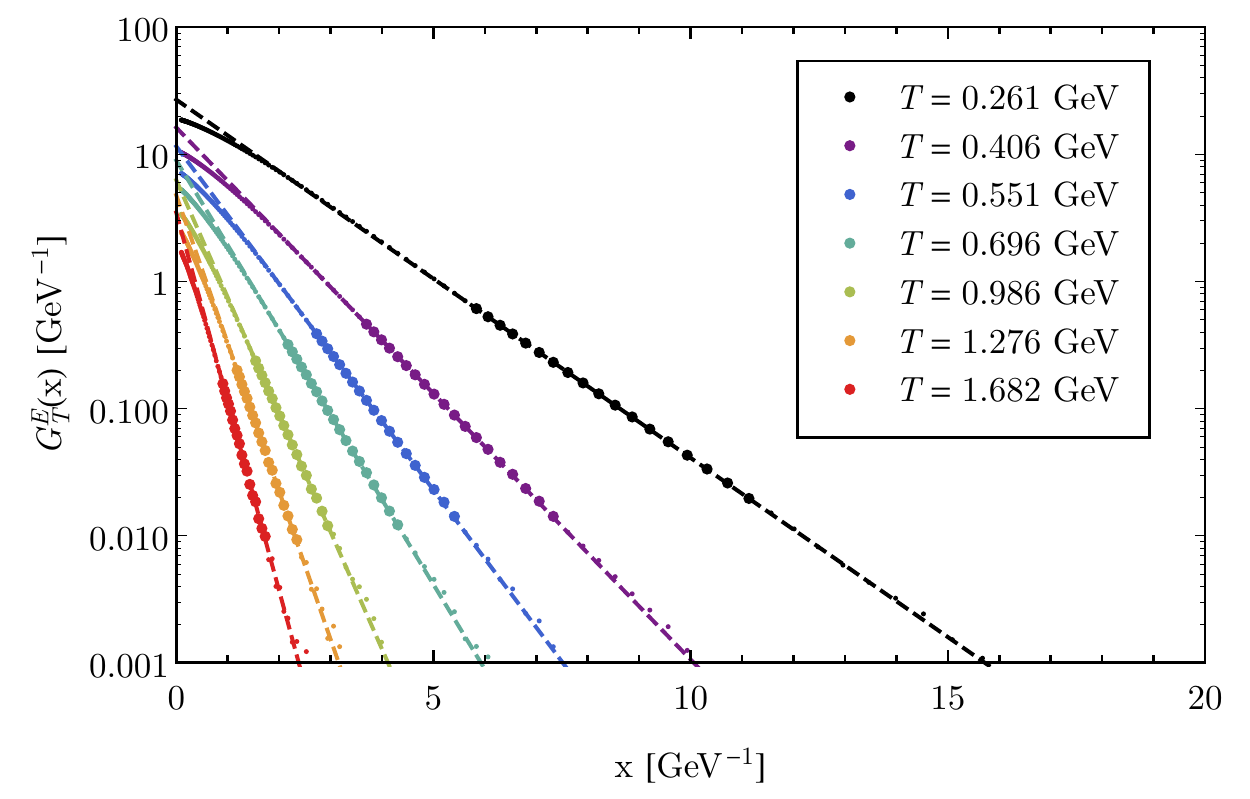}}{
				\caption{$\GE(x)$ for high temperatures.}
				\label{fig:DebyeMassFit:HighT}
			}
		\end{subfloatrow}
	}{
		\caption{
			Exponential tail of the Fourier transformed
			electric propagator, $\GE(x)\,$, see
			\eq{eq:SpatialProp}. The dashed lines are fits of
			\eq{eq:DecayTapp} with $c_a=0\,$, \ie \eq{eq:DecayT},
			to the large points.  Small points show
			$\GE(x)$ beyond the fit regions.
			The fitted screening masses as a function of temperature are shown in \Fig{fig:DebyeMass}.
			\myhfill
		}
		\label{fig:DebyeMassFit}
	}
\end{figure*}

\section{Gluon mass parameter}
\label{app:UVmass}

The gluon mass parameter,
$m_\Lambda^2\propto\alpha(\Lambda)\,\Lambda^2\,$, is fixed at the
cutoff scale $k=\Lambda\,$, which is far bigger than the temperature
scale, $\Lambda\gg \twopiT\,$. As discussed in
\Sec{sec:Method:Renormalisation}, this parameter is determined by the
modified Slavnov-Taylor identity that is difficult to solve
numerically with the required quadratic precision. In addition, a
truncation to the combined system of mSTI and flow has to be
self-consistent to quadratic precision.  At
vanishing temperature, we have therefore determined the gluon mass
term by requiring a solution of the scaling type.

Temperature effects are suppressed exponentially for the used
regulators, see \cite{Fister:2015eca} and
\Sec{sec:Method:FRG}. Hence, the initial conditions for the flow at
$k=\Lambda$ converge exponentially to that at vanishing
temperature. However, in the present scaling solution the initial
conditions compensate for the violation of the modified BRST-symmetry
during the flow, and in particular at low cutoff scales. Therefore, we
expect a temperature-dependent change of the initial conditions for
compensating temperature-dependent truncation artefacts at low scales.
Keeping this in mind, we extend the BRST-consistent fine-tuning of the
initial conditions to finite temperature, 
\begin{align}
	\label{eq:TemperatureMassTuning}
	\Gamma^{(2),\text{M,raw}}_{AA,k=\Lambda_T}(p) =
	\Gamma^{(2),\text{E,raw}}_{AA,k=\Lambda_T}(p) = 
	\Gamma^{(2),T=0}_{AA,k=\Lambda_T}(p) + \Delta m^2_T\,. 
\end{align}
The temperature-dependent part of the gluon mass parameter
$\Delta m^2_T$ is fixed such that we obtain infrared
scaling in the purely magnetic sector, see
\Sec{sec:Method:Renormalisation}. 
Its sole purpose is to adjust the
modified BRST symmetry as in the $T=0$ case. 
Requiring scaling fixes $\Delta m^2_T$ uniquely.
While adjusting the correct infrared behaviour,
this introduces truncation artefacts in the UV. The RG-relevant part
of the temperature-dependence at large momentum has to vanish
identically. It is uniquely removed with
\begin{align}
	\label{eq:GluonPropagotorCorrection}
	\Gamma^{(2),\text{M/E}}_{AA,k=0} =
		\Gamma^{(2),\text{M/E,raw}}_{AA,\,k=0} -
		\left(
			\Gamma^{(2),\text{M/E,raw}}_{AA,\,k=k_T} -
			\Gamma^{(2),T=0}_{AA,\,k=k_T}
		\right)\,. 
\end{align}
Here, $k_T\approx \fourpiT \leq \Lambda_T$ is the scale above which
temperature effects are virtually absent, \ie
\begin{align}
	\Gamma^{(2),\text{M,raw}}_{AA,\,k\geq k_T}=\Gamma^{(2),\text{E,raw}}_{AA,\,k\geq k_T}\, .
\end{align}
Note that \eq{eq:GluonPropagotorCorrection} keeps the physical
temperature-dependent polynomially suppressed large momentum
corrections, see
\cite{Fister:2011uw}. Equation \eq{eq:GluonPropagotorCorrection} removes in
particular $\Delta m^2_T\,$ from the final result.  Moreover, the
mass correction \eq{eq:TemperatureMassTuning} leads to modifications
of the flows due to the back coupling of the changed gluon mass
parameter.  Consequently, the subtraction
\eq{eq:GluonPropagotorCorrection} removes back-coupling artefacts that
are accumulated during the integration of the flow from $\Lambda_T$ to
$k_T$. In the case $\Lambda_T=k_T$, no back-coupling artefacts are
created at scales larger than $k_T$ and the correction becomes
trivial,
\begin{align}
	\label{eq:GluonPropagotorCorrectionReduced}
	\Gamma^{(2),\text{M/E,raw}}_{AA,\,k=k_T} - \Gamma^{(2),T=0}_{AA,\,k=k_T}=\Delta m^2_T\,.
\end{align}
We demonstrate in \Fig{fig:StartScale} that neither the raw nor the 
final gluon propagators, obtained with
\eq{eq:GluonPropagotorCorrection}, depend on the initial cutoff scale
$\Lambda_T\,$.  Thus, $\Lambda_T=k_T$ is the numerically least
demanding and most stable choice that includes all thermal
fluctuations. Note that \eq{eq:TemperatureMassTuning} modifies the
magnetic and electric propagators identically.  Thus, the electric
mass is an observable at vanishing cutoff.

In order to assess the effect of the temperature-dependent tuning of
the gluon mass parameter \eq{eq:TemperatureMassTuning}, we compare the
raw with the final propagators in \Fig{fig:EffectOfSubtraction}. We
observe a sizeable influence of the correction on the final result.
We plot the raw and the final magnetic gluon propagator each
normalised by the magnetic lattice propagator in
\Fig{fig:SubtractionLatticeComparison}.  Since the lattice data have
to be renormalised for each temperature separately, agreement is
always found at the corresponding momentum scale, see
\App{app:ScaleSettingAndRenormalisation}.  At lower scales, the raw
propagator quickly deviates from the lattice results.  Contrarily, the
final propagator shows better agreement, which we interpret as support
for our subtraction procedure \eq{eq:GluonPropagotorCorrection}.

\section{Scale setting and renormalisation}
\label{app:ScaleSettingAndRenormalisation}

We set the scale by rescaling our internal units such that the bump
position of the gluon propagator dressing lies at
$p_\text{max}\equiv\SI{0.955}{\GeV}$, and thus coincides with the bump
position of the vacuum lattice results from \cite{Sternbeck:2006cg},
see \Fig{fig:RegulatorIndependence}.

The temperatures of the $SU(2)$ lattice results from
\cite{Maas:2011ez,Maas:PC} are given in terms of the critical
temperature.  In order to compare, we use
$T_c^{SU(2)}=0.7091\,\sqrt{\sigma}=\SI{312}{\MeV}$
\cite{Lucini:2003zr} to convert the temperature into units of
\si{\GeV}, where the string tension $\sigma$ is given by
$\sigma=\SI{0.440}{\GeV\squared}$.  These lattice results need to be
renormalised for each temperature separately.  We determine the
temperature-dependent renormalisation constants by fitting all lattice
points above $p\geq \max\left(\twopiT,\,\SI{1}{\GeV}\right)$ to our
results.

The $SU(3)$ lattice results from \cite{Silva:2013maa} do not include
the vacuum case $T=0\,$.  Therefore, we allow for a scale mismatch by
introducing a temperature-independent relative scale factor $r_s\,$,
in addition to the temperature-independent wave function
renormalisation constant $z_L\,$.  We determine $r_s$ and $z_L$ by
fitting the magnetic gluon dressing function,
$1/\left(z_L\, \ZAM(r_s T\,, r_s p)\right)\,$, simultaneously for all
temperatures to all lattice points above $p\geq\SI{0.5}{\GeV}$.
Subsequently, we use $r_s$ and $z_L$ to rescale the magnetic as well
as the electric lattice propagators to our data.  We find the relative
scale mismatch $r_s-1$ to be small, of the order of
$\SI{2}{\percent}$. The temperatures in \cite{Silva:2013maa} are
given in units of \si{\GeV}.  In order to simplify the discussion, we
convert the temperatures into units of the critical temperature, using
their value for the $SU(3)$ phase transition temperature,
$T_c=\SI{270}{\MeV}$.

\section{Screening mass}
\label{app:Screening}

In this appendix we describe the
extraction of the screening masses shown in \Sec{sec:Setup:DebyeMass}
and provide the propagators in position space, see \Fig{fig:DebyeMassFit}.
In general, the thermal propagators at large distances show a
combination of an exponential and an algebraic decay,
\begin{align}
	\label{eq:DecayTapp}
	\lim_{x\to \infty} \; G^{\text{E}}_{T}(x) = 
		c_a \, x^{1-4\kappa} + c_e \, \exp\left(-m_s \, x\right)\,,
\end{align}
where $G^{\text{E}}_{T}(x)$ is the Fourier transformed zero mode of the electric
gluon propagator \eq{eq:SpatialProp},
see \Sec{sec:Setup:DebyeMass}.
The algebraic decay originates from
the infrared scaling at vanishing temperature with $1-4\kappa$ being
the scaling exponent in position space, \cf also
\eq{eq:GeneralScaling}.  At low enough temperatures $T\ll T_c\,$, we
see remnants of this zero-temperature algebraic part. At higher
temperatures we have $c_a =0$.

In \Fig{fig:DebyeMassFit}, we show $\GE(x)$ for low (left
panel) and high (right panel) temperatures. For all shown temperatures,
the exponential decay is apparent as linear regime.
For even lower temperatures, the linear regime shrinks considerably
due to the algebraic decay, and we need a higher
numerical precision for extracting the sub-leading exponential
decay. This is clearly seen in the lowest temperature shown,
$T=\SI{0.102}{\GeV}$.

\bibliography{../../bib_master}

\end{document}